\documentclass[reprint,
	10pt, notitlepage,
    floats, floatfix,
	amsmath, amssymb, amsfonts, 
	superscriptaddress,
	showpacs, showkeys, aps,
    floatfix, nofootinbib,prd
]{revtex4-2}
\usepackage{amsmath, amssymb, amsfonts}
\usepackage[utf8]{inputenc} 
\usepackage[usenames,dvipsnames]{xcolor}
\xdefinecolor{mylinkcolor}{rgb}{0,0,0.5}
\usepackage[bookmarksnumbered, bookmarksopen, bookmarksopenlevel=2,breaklinks=true,colorlinks=true, filecolor=mylinkcolor,citecolor=mylinkcolor,linkcolor=mylinkcolor,urlcolor=mylinkcolor,menucolor=mylinkcolor]{hyperref}
\usepackage{cancel, aasmacros, graphicx, dcolumn, bm, lipsum, xspace, array, verbatim, color, lineno, bookmark}
\usepackage[normalem]{ulem}
\usepackage[inline]{enumitem}
\usepackage{orcidlink}
\usepackage{cleveref}

\newcommand{\Ins}{\affiliation{Dipartimento di Scienza e Alta Tecnologia, Università dell’Insubria, via Valleggio 11, I-22100 Como, Italy}}
\newcommand{\Bic}{\affiliation{Dipartimento di Fisica “G. Occhialini”, Università degli Studi di Milano-Bicocca, Piazza della Scienza 3, 20126 Milano, Italy}}
\newcommand{\Infn}{\affiliation{INFN, Sezione di Milano-Bicocca, Piazza della Scienza 3, 20126 Milano, Italy}}
\newcommand{\bham}{\affiliation{Institute for Gravitational Wave Astronomy \& School of Physics and
Astronomy, University of Birmingham, Birmingham, B15 2TT, UK}}
\newcommand{\inaf}{\affiliation{INAF – Osservatorio Astronomico di Brera, via E. Bianchi
46, 23807, Merate, Italy}}
\newcommand{\mpi}{\affiliation{Max-Planck-Institut f{\"u}r Astrophysik, Karl-Schwarzschild-Stra\ss e 1, 85741 Garching, Germany}}

\newcommand{\dd}{\mathrm{d}}

\begin{document}
\preprint{APS/123-QED}
\title{Cyclostationary signals in LISA: a practical application to Milky Way satellites}

\author{Federico~Pozzoli\orcidlink{0009-0009-6265-584X}} \email{fpozzoli@uninsubria.it} \Ins
\author{Riccardo~Buscicchio\orcidlink{0000-0002-7387-6754}}\Bic \Infn \bham  
\author{Antoine~Klein\orcidlink{0000-0001-5438-9152
}} \bham 
\author{Valeriya~Korol\orcidlink{0000-0002-6725-5935
}} \mpi \bham 
\author{Alberto Sesana\orcidlink{0000-0003-4961-1606}}\Bic \Infn \inaf
\author{Francesco Haardt\orcidlink{0000-0003-3291-3704}}\Ins \Infn \inaf

\begin{abstract}
One of the primary sources of gravitational waves (GWs) anticipated to be detected by the Laser Interferometer Space Antenna (LISA) are Galactic double white dwarf binaries (DWDs). However, most of these binaries will be unresolved, and their GWs will overlap incoherently, creating a stochastic noise known as the Galactic foreground. Similarly, the population of unresolved systems in the Milky Way's (MW) satellites is expected to contribute to a stochastic gravitational wave background (SGWB). Due to their anisotropy and the annual motion of the LISA constellation, both the Galactic foreground and the satellite SGWB fall into the category of cyclostationary processes. Leveraging this property, we develop a purely frequency-based method to study LISA's capability to detect the MW foreground and SGWBs from the most promising MW satellites.
We analyze both mock data generated by an astrophysically motivated SGWB spectrum, and realistic ones from a DWD population generated via binary population synthesis.
We are able to recover or put constrains on the candidate foregrounds, reconstructing ---in the presence of noise uncertainties--- their sky distribution and spectrum.
Our findings highlight the significance of the interplay between the astrophysical spectrum and LISA's sensitivity to detect the satellites' SGWB. Considering an astrophysically motivated prior on the satellite positions improves their detectability, which becomes otherwise challenging in the presence of the Galactic foreground. Furthermore, we explore the potential to observe a hypothetical satellite located behind the Galactic disk. Our results suggest that a Large Magellanic Cloud-like satellite could indeed be observable by LISA.
\end{abstract}

\maketitle

\section{Introduction}
The Laser Interferometer Space Antenna (LISA)~\cite{2017arXiv170200786A} is a groundbreaking mission for the detection of Gravitational Waves (GW) from space. It will be sensitive to the milli-Hz (mHz) frequency range, and during its nominal $4$ years of observation will resolve tens of thousands of astrophysical sources of different nature, from supermassive black hole binaries colliding in the young Universe, to stellar compact objects within our solar neighborhood \cite{2024arXiv240207571C}.


Besides individual sources, the Universe is pervaded by a stochastic gravitational wave background (SGWB), which arises from the incoherent superposition of GWs originating from numerous unresolved or weak sources.
At nano-Hz frequencies, the first evidence for an unresolved GW signal of astrophysical and/or cosmological origin~\cite{2023ApJ...951L..11A,2024A&A...685A..94E} has recently been reported by the European pulsar timing array~\cite{2023A&A...678A..50E}, NANOGrav~\cite{2023ApJ...951L...8A}, the Parkes Pulsar timing array~\cite{2023ApJ...951L...6R} and the Chinese pulsar timing array~\cite{2023RAA....23g5024X}. 
The SGWB in LISA may come from both cosmological and astrophysical sources \cite{2016JCAP...12..026B,2016JCAP...04..001C,2020PhRvD.102j3023B,2023PhRvD.108j3039P,2023JCAP...08..034B,2024A&A...683A.139S}. The Galactic foreground~\cite{2009CQGra..26i4030N} from double white dwarf binaries (DWDs) is expected to dominate, particularly in the $0.5-3$ mHz range. 
The bulk of such foreground arises from systems located toward the Galactic center, hence it is expected to be highly anisotropic. The corresponding distribution of unresolved sources, combined with the annual motion of LISA, results in a modulated signal in the time domain, yielding a cyclostationary stochastic process. 
The relevance of such statistical property is twofold: it allows to better characterize the Milky Way (MW) foreground, and it facilitates the study and identification of other signals. 
It has been studied in previous works~\cite{2024JCAP...06..055M, 2022ApJ...940...10D}, although it has typically been modeled using empirical formulas. 
However, Ref.~\cite{Buscicchio2024} has  introduced a physically based, parameterized template that accounts for a specific distribution in the sky. The authors also demonstrated that between $2-10 \rm mHz$, the MW foreground exhibits a significant deviation from Gaussianity. 
This property will be central in accurately characterizing the confusion noise.

Recent simulations indicate that besides the MW, other nearby dwarf galaxies such as the Large Magellanic Cloud (LMC), Small Magellanic Cloud (SMC), Sagittarius, Sculptor, and Fornax also host DWDs that are potentially resolvable as individual sources by LISA~\cite{2020ApJ...894L..15R,2023MNRAS.521.1088K}.
The capacity to resolve DWDs depends on factors such as the mass, distance, and star formation history of the dwarf galaxy, as noted in Ref.~\cite{2020A&A...638A.153K}. 
Nonetheless, similarly to the MW, most DWDs within dwarf galaxies are expected to remain unresolved. 
Therefore, it is likely that the undetectable DWDs might collectively contribute to a SGWB detectable by instruments like LISA. 
Alike the Galactic foreground, SGWB from DWD binaries in satellites will also be relevant to better understand the properties of the underlying astrophysical population, such as the total stellar mass, metallicity, and star formation history ~\cite{2020ApJ...901....4B,2023MNRAS.519.2552G}.
An extensive study of the SGWB detection from LMC has been presented in Ref.~\cite{2024MNRAS.531.2642R}, where the anisotropy is modelled through spherical harmonics decomposition, already employed to study anisotropies coming from cosmological processes. 

In this paper, we introduce a method to characterize a cyclostationary SGWB in LISA. This approach allows us to study the detectability of SGWB from satellite galaxies or nearby galaxies and to infer its parameters. 

This paper is organized as follows. We present in Sec.~\ref{sec:cyc} a detailed construction of our frequency domain Gaussian cyclostationary model.
In Sec.~\ref{sec:lisa-data} we further describe the underlying assumptions of our analysis.
Then in Sec.~\ref{sec:res}, we discuss our results. 
First, we analyze the signal satellite considering only instrumental noise. Then, we incorporate the Galactic foreground and assess its impact on the reconstruction of SGWB satellites. To validate our algorithm, we test it on both mock and realistic datasets. Furthermore, to improve the detectability of these signals, we explore the use of informed priors derived from electromagnetic observations.
We also investigate the capacity of our approach to resolve an LMC-like satellite placed behind the MW disk. 
We conclude and discuss possible future outlooks in Sec.~\ref{sec:conc}.

\section{Cyclostationary signals in LISA}\label{sec:cyc}

Cyclostationary processes are stochastic processes whose statistical properties are periodic in time. 
In this work, we focus on those exhibiting periodicity in their second-order statistics (e.g., the autocorrelation function)~\cite{VANKAMPEN200752}. 
In particular, a continuous stochastic process $X(t)$ having finite second-order moments is said to be wide-sense cyclostationary with period $T$ if the expectation values
\begin{align}
    &E\left[X(t)\right] = m(t) = m(t +T) \\
    &E\left[X(t^\prime)X(t)\right] = \Sigma(t^\prime,t) = \Sigma(t^\prime + T,t + T)  
\end{align}
are periodic functions with period $T$, for $(t^\prime,t) \in \mathbb{R}^2$. 
Defining $\tau_d= t^\prime - t$, then $\Sigma(t^\prime,t)$ can be equivalently represented as a Fourier series:
\begin{align}
    \label{eq:cyclic_corr}
    \Sigma(t^\prime,t)= \sum_{n=-\infty}^{+\infty}B_n(t^\prime -t)e^{-2\pi {\rm i}nt/T},
\end{align}
where $B_n(\tau_d)$ is
\begin{equation}
    \label{eq:cyclic_auto}
    B_n(\tau_d) = \frac{1}{T}\int_0^T \dd t \Sigma(t+\tau_d,t)e^{-2\pi {\rm i} nt/T}.
\end{equation}
The Fourier transform of Eq.\eqref{eq:cyclic_auto} corresponds to the so-called \textit{cyclic spectrum}. 
We contextualize the above formalism to populations of unresolved DWDs in the MW and its satellites, as seen by LISA.
For an ensemble of $N$ DWDs, the total signal $s(t)$ is given by 
\begin{equation}
    s(t) = \sum_{i=0}^N X_{\rm GW}(t, \bf{\Theta})
\end{equation}
where $\bf{\Theta}$ is the vector of parameters characterizing the GW signal $X_{\rm GW}$.
We list them here for completeness: the initial phase of the signal $\phi_0$, the inclination angle $\iota$ of the binary system's angular momentum relative to the line of sight, the polarization angle $\psi$, the distance $D$ to the binary, the ecliptic coordinates $(\lambda,\beta)$ describing the source locations in the solar system barycenter frame (SSB), the chirp masses $\mathcal{M}_c$, and the angular frequencies $\omega_s$. $X_{\rm GW}$ is the signal in one of the three LISA noise-orthogonal channel~\cite{2005LRR.....8....4T}. 
We will consider only sources in their early inspiral, such that in the low-frequency approximation, each monochromatic signal in $X_{\rm GW}$ is expanded as
\begin{align}
\label{eq:xgw}
    X_{\rm GW}(t) = &2C(t)(1+\cos^2\iota)\cos(\omega_s t + \phi_0)A + \nonumber \\
    & + 4S(t)\cos\iota\sin(\omega_s t + \phi_0)A  
\end{align}
where $C(t)$ and $S(t)$ depend on the motion of the LISA satellites constellation, and 
\begin{align}
    A &= \frac{(G\mathcal{M}_c)^{5/3}}{c^4D}\left(\pi f_s\right)^{2/3},\\
    \mathcal{M}_c &= \frac{(m_1m_2)^{3/5}}{(m_1+m_2)^{1/5}}.
 \end{align}
Here, $A$ is the GW amplitude and $\mathcal{M}_c$ is the chirp mass as function of the binary component masses $m_1$ and $m_2$. $G$ and $c$ are respectively the gravitational constant and the light speed.
The largest fraction of MW DWDs can be roughly approximated to share the same distance $D$ (this is even more true for MW satellites). Thus, to compute the autocorrelation function $\Sigma(t^\prime,t)$, we assume that the population distribution $p(\bf{\Theta})$ is separable into five distributions, $p_1(\phi_0)$, $p_2(\iota)$, $p_3(\psi)$, $p_4(\beta,\lambda)$, and $p_5(\mathcal{M}_c,\omega_s)$.
Specifically, we assume probability distributions for $\psi$ ($\phi_0$, $\cos\iota$) uniform in $\left[0,2\pi\right]$ ($\left[0,2\pi\right]$, $\left[-1,1\right]$). 
In what follows, we refer for brevity to whole DWD populations in MW satellites as ``sources''.
Then, we compute the autocorrelation function as
\begin{align}
\label{eq:autocorr}
    \Sigma(t^\prime,t) &=N\int^{2\pi}_0 \!\!\!\!\!\dd\phi_0\int^{2\pi}_0 \!\!\!\!\!\dd\psi \int^{1}_{-1} \!\!\!\!\dd\iota \int_{V_4}\!\!\!\dd V_4 \frac{p_4}{16\pi^2} \int_{V_5}\!\!\!\dd V_5 p_5 \times\nonumber \\
    &\times X_{\rm GW}(t^\prime)X_{\rm GW}(t),
\end{align}
where $V_4$ ($V_5$) is the differential volumes of the ecliptic latitude and longitude $\beta,\lambda$ (chirp mass, orbital angular frequency $\mathcal{M}_c,\omega_s$) parameter space, respectively.
By inserting Eq.\eqref{eq:xgw} in Eq.\eqref{eq:autocorr} and performing the integral over $\psi$, $\cos\iota$ and $\phi_0$, we obtain:
\begin{align}
   \Sigma(t^\prime,t) &=\frac{N}{8}\int_{V_4}\dd V_4 p_4\int_{V_5}\dd V_5p_5 A^2 \cos\left(\omega_s(t^\prime-t)\right)\times \nonumber\\
   &\times \left[\frac{56}{15}C(t^\prime)C(t) + \frac{8}{3}S(t^\prime)S(t)\right].  
\end{align}
Introducing the additional variable $\tau_s = \frac{t +t ^\prime}{2}$, we define $B(\tau_s,\tau_d)= \Sigma(\tau_s + \tau_d/2,\tau_s - \tau_d/2)$. Therefore, the autocorrelation $B(\tau_s,\tau_d)$ is explicitly periodic in $\tau_s$ for any fixed $\tau_d$. Thus, similarly to Eq.~\eqref{eq:cyclic_corr} we decompose it in Fourier series
\begin{align}\label{eq:autocorr2}
    B(\tau_s,\tau_d) & = \frac{N}{8}\int_{V_5}\dd V_5p_5 \sum_{n=-8}^{n=8}B_n(\tau_d)e^{2\pi i\frac{n \tau_s}{T}}\times \nonumber\\
    & \times A^2\cos(\omega_s\tau_d) .
\end{align}
Note that the limitation to $|n| \leq 8$ is a known result for the Fourier series decomposition of product of oscillating functions at harmonic frequencies (see e.g.~\cite{1997MNRAS.291..149G}).

\begin{figure*}[t]
    \centering    \includegraphics[width=0.45\textwidth]
    {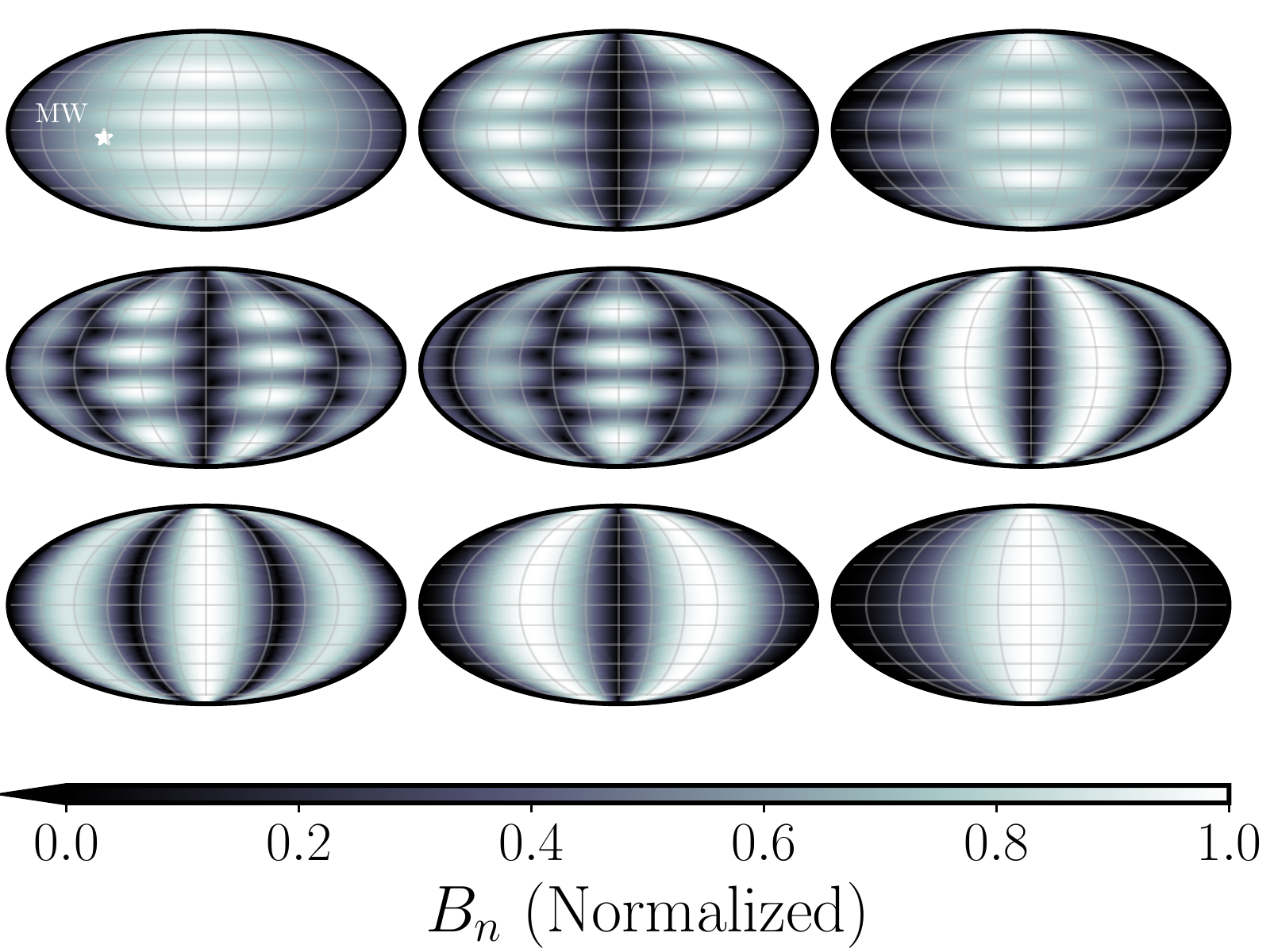}
    \hspace{0.55cm}   
    \includegraphics[width=0.45\textwidth]
    {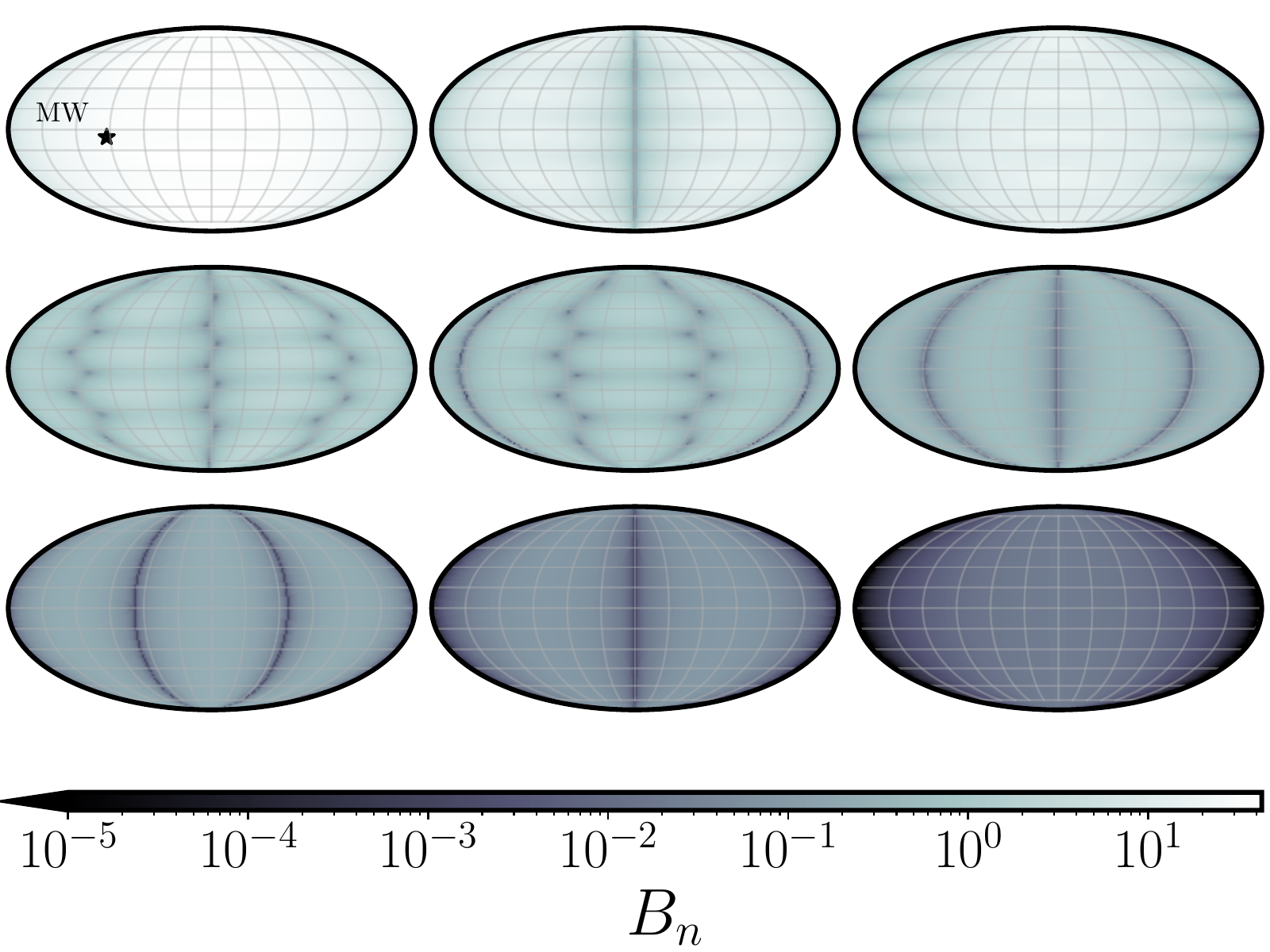}
    \caption{Harmonic decomposition of LISA modulation. Each point in the sky corresponds to a source with the angular size of the LMC ($\sigma \approx 0.015$ rad). In each panel, the plots are associated with harmonics in increasing order of $n$ from top left ($n=0$) to bottom right ($n=8$). The projections are expressed in ecliptic coordinates. Left Panel: each heatmap represents the magnitude of the Fourier Coefficient normalized to the relative maximum value in each plot. Right Panel: the absolute magnitude of the Fourier coefficient is shown in logarithmic scale. }
    \label{fig:harmonic}
\end{figure*} 

Further assuming that $p_5(\mathcal{M}_c,\omega_s)$ is only mildly frequency dependent -- or equivalently that $\tau_d$ is such that $\tau_d/T \ll 1$-- we can approximate $B_n(\tau_d) = B_n(\tau_d = 0)$, as in Ref.~\cite{2005PhRvD..71l2003E}.
Therefore, we will refer to it in what follows as $B_n$.
We express the autocorrelation function in Eq.~\eqref{eq:autocorr2} as product of two functions
\begin{equation}
   B(\tau_s,\tau_d) = R(\tau_s)\mathcal{C}(\tau_d)
\end{equation}
where
   \begin{align}
   R(\tau_s) &= \sum_{n=-8}^{n=8}B_n e^{2\pi i\frac{n \tau_s}{T}} \\
   \mathcal{C}(\tau_d) &= \frac{N}{8}\int_{V_5}\dd V_5p_5A^2\cos(\omega_s\tau_d)\label{eq:autocorr3}
\end{align}
By Fourier transforming over $\tau_d$ and $\tau_s$, we obtain
\begin{align}
\label{eq:cov_freq}
C(f,f^\prime)  =\sum_{n=-8}^{n=8}B_nS_h\left(\frac{f^\prime +f}{2}\right)\delta\left(f-f^\prime + \frac{n}{T}\right),
\end{align}
where $\delta$ is the Dirac delta function.
In a discretized form (which is the natural domain for an observed digital timeseries), $C(f,f^\prime)$ corresponds to a band matrix $C_{ij}$ with $2n+1=17$ non-zero diagonals spaced $|i-j|= T_{\rm obs}/T$ apart from each other. 
Here, the periodicity of the correlation function is $T$ and the frequency resolution is limited by the observation time (i.e. the LISA mission duration) $T_{\rm obs}$.
Each diagonal is proportional to time-series one-sided power spectral density $S_h$, which depends on the intrinsic astrophysical properties of the source population, defined as 
\begin{equation}
S_h(f)=\lim_{T\rightarrow +\infty}\frac{1}{2T} \left\langle \left| \int_{-T}^{+T}{\rm d}t\ x(t) e^{- {\rm i} 2\pi f t}\right|^2\right\rangle .
    \label{eq:stoch_proc_spectrum}
\end{equation}
where $\langle \cdot \rangle$ denote averaging over an ensemble of realizations.

The proportionality constants $B_n$ are the $n$-th harmonic Fourier coefficients of the time-domain signal modulation, and they depend solely on the extrinsic properties of the source population, e.g. the sky distribution. 
This is a crucial advantage of our approach, as it directly relates the signal modulation to the source distribution.
As compared to other approaches in literature, it does not require either pixelation or spherical harmonics decomposition.
Once a suitable parameterization for $C_{ij}$ is introduced for a specific detector, it allows for a straightforward interpretation of the inference results. In Sec.~\ref{sec:lisa_mod} we present such a model for LISA.

\subsection{LISA Modulation}\label{sec:lisa_mod}
Following the detailed mathematical derivation in Ref.~\cite{Buscicchio2024}, we construct the LISA modulation for a bivariate Gaussian distribution in the sky-coordinates. 
Therein, the contribution of different harmonics to the overall time domain modulation are listed.
In particular, the Fourier coefficient $B_n$, are described by five independent parameters: 
the ecliptic coordinates of the distribution expectation value $(\sin \beta, \lambda)$, and the two principal component axes variances $(\sigma_1, \sigma_2)$, along with their rotation angle $\psi$ with respect to the coordinate system. 
Given the modest angular size of MW satellites, we choose to enforce the constraint $\sigma_1=\sigma_2$, which reduces the number of free parameters to four, with $\psi$ being completely degenerate. 
Therefore, the only parameters left are  $\sigma \equiv  \sigma_1, \sin \beta, \lambda $, and our inference model is therefore specified by $R(\tau_s; \sin \beta, \lambda,  \sigma)$.

In what follows, we explore the satellite modulation as a function of its free parameters.
In Fig.~\ref{fig:harmonic}, we show the magnitude of the Fourier coefficients across the sky, each panel corresponding to a single harmonic, for a reference $\sigma\approx 0.015$ rad. 
We highlight that due to the modulation being a real function of time, $B_{-n}=B^*_{n}$.
The zero-th mode, which corresponds to the main diagonal in the covariance matrix, is dominant compared to the others.
For a given sky position, the higher-order modes are increasingly subdominant. 
The odd (even) harmonics share common maxima (minima) along the directions $(\sin \beta,0)$ , although we point out that the specific value depends crucially on the chosen initial LISA position along its orbit.
Overall, far from the ecliptic plane, the LISA modulation becomes less relevant: if a source is located close to the ecliptic poles, the Doppler effect induced by the LISA orbital motion is negligible. 
Conversely, the ecliptic longitude appears to have a less significant influence on the modulation.

\begin{figure}
    \centering    \includegraphics[width=1.0\columnwidth]{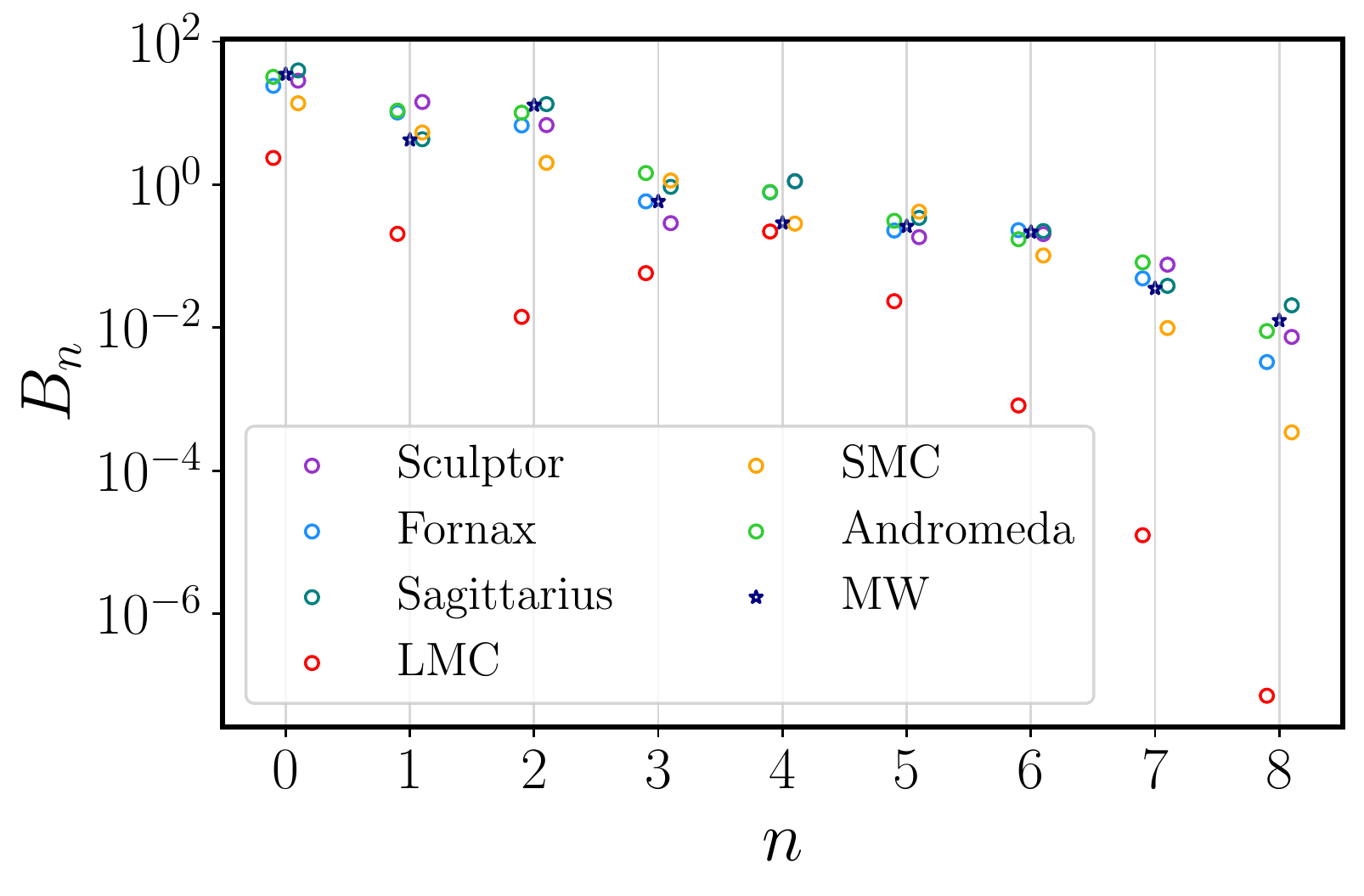}
    \caption{Magnitude of Fourier coefficients in logarithmic scale at various harmonics for the MW, Andromeda and the set of satellites considered in this study. Higher harmonics are generally less significant. The values for the LMC are lower compared to other sources due to its proximity to the ecliptic South Pole. Individual source parameters are listed in Table~\ref{tab:par}.\label{fig:harmonic_sigma}}
\end{figure}

In Fig.~\ref{fig:harmonic_sigma}, we additionally show the magnitude of each Fourier coefficient for a set of known MW satellites (together with Andromeda), whose parameters are listed in Table~\ref{tab:par} (for more details, see Ref.~\cite{2020ApJ...894L..15R} and references therein). These are the sources we will consider in this study.
We observe that the size of a source may also play a role on its modulation. In particular, smaller objects result in larger coefficients.
However, this effect is less significant compared to that of the ecliptic latitude and longitude.
Sagittarius and the MW are relatively close to each other in the sky, but some $B_n$s of the former are larger than those of the latter (e.g. $n=4$).
On the contrary LMC, located near the ecliptic South Pole, yields much weaker modulation than the MW even if it has a comparable angular size to Sagittarius.

\begin{figure*}
    \includegraphics[width=\textwidth]{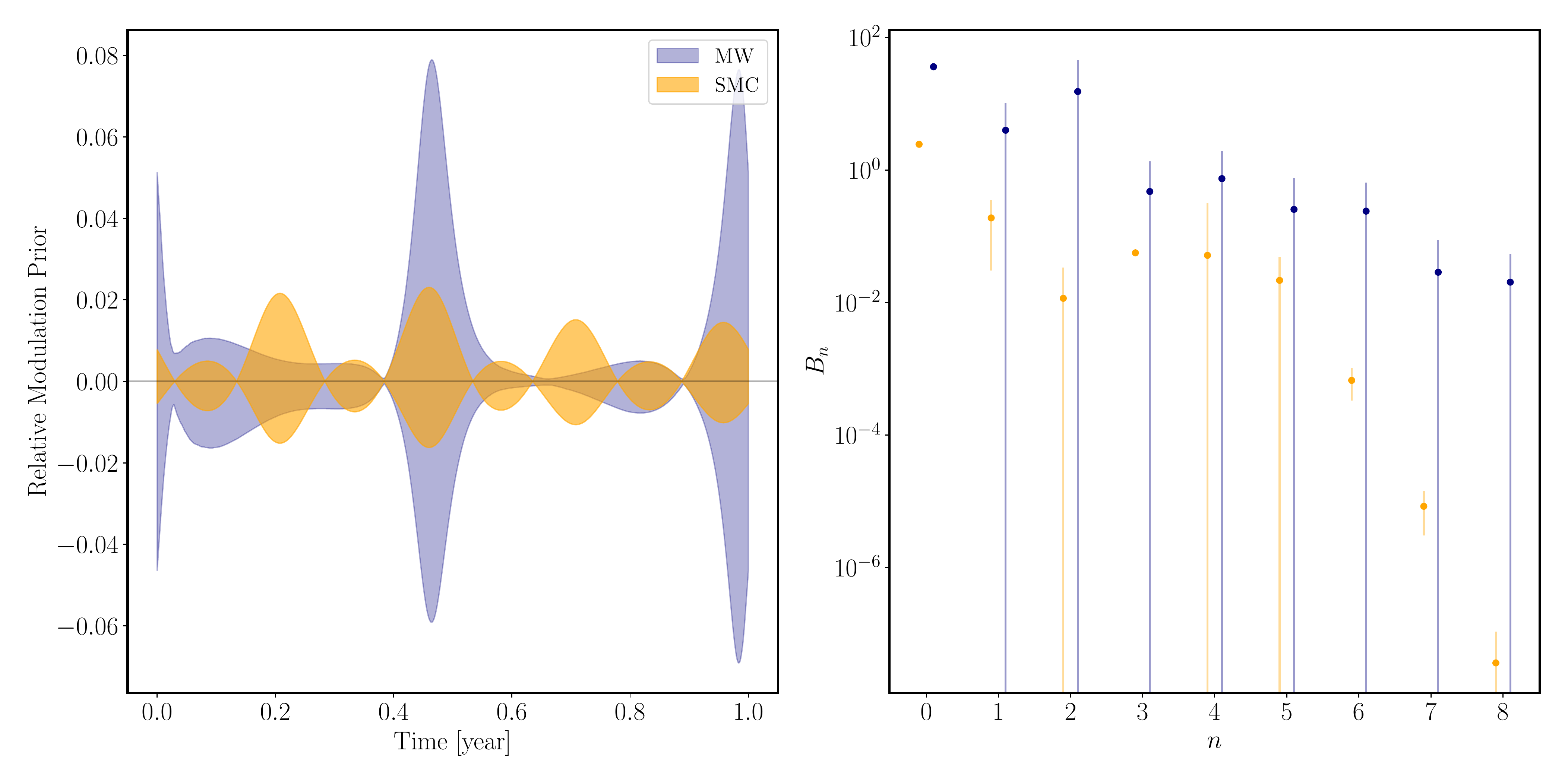}
    \caption{The Impact of varying $\sigma$: Each panel shows the modulation calculated by varying $\sigma$ by $\pm 50\%$ of its original value, while keeping the true position of the relative object fixed. Left Panel: The $90\%$ confidence interval of the relative time modulation. Right Panel: The $90\%$ confidence intervals for different harmonic modes. Overall, the SMC modulation shows less variability compared to the MW when exploring the $\sigma$ prior. 
    }
    \label{fig:prior_harm}
\end{figure*}

We further investigate the impact of the source angular size in Fig.~\ref{fig:prior_harm}.
In the left panel, 
we show the relative uncertainties in the time modulation for the SMC and the MW, varying their $\sigma$ by $\pm 50\%$. 
For the Galactic case, relative deviations reach nearly $10\%$ during a LISA orbit, as opposed to a maximum of $2\%$ for SMC. 
Similarly, in the right panel of Fig.~\ref{fig:prior_harm}, the same variability is illustrated in frequency domain. Ranges for each MW modulation coefficient are much larger than the corresponding ones for the SMC. 
Hence, the modulation phenomenology is only mildly influenced by the satellite angular size, thus only marginally affecting the SGWB detection.
A more detailed discussion is provided in Sec.~\ref{sec:res}.

\subsection{Astrophysical Spectrum}
\begin{figure*}
    \centering
    \includegraphics[width=\textwidth]{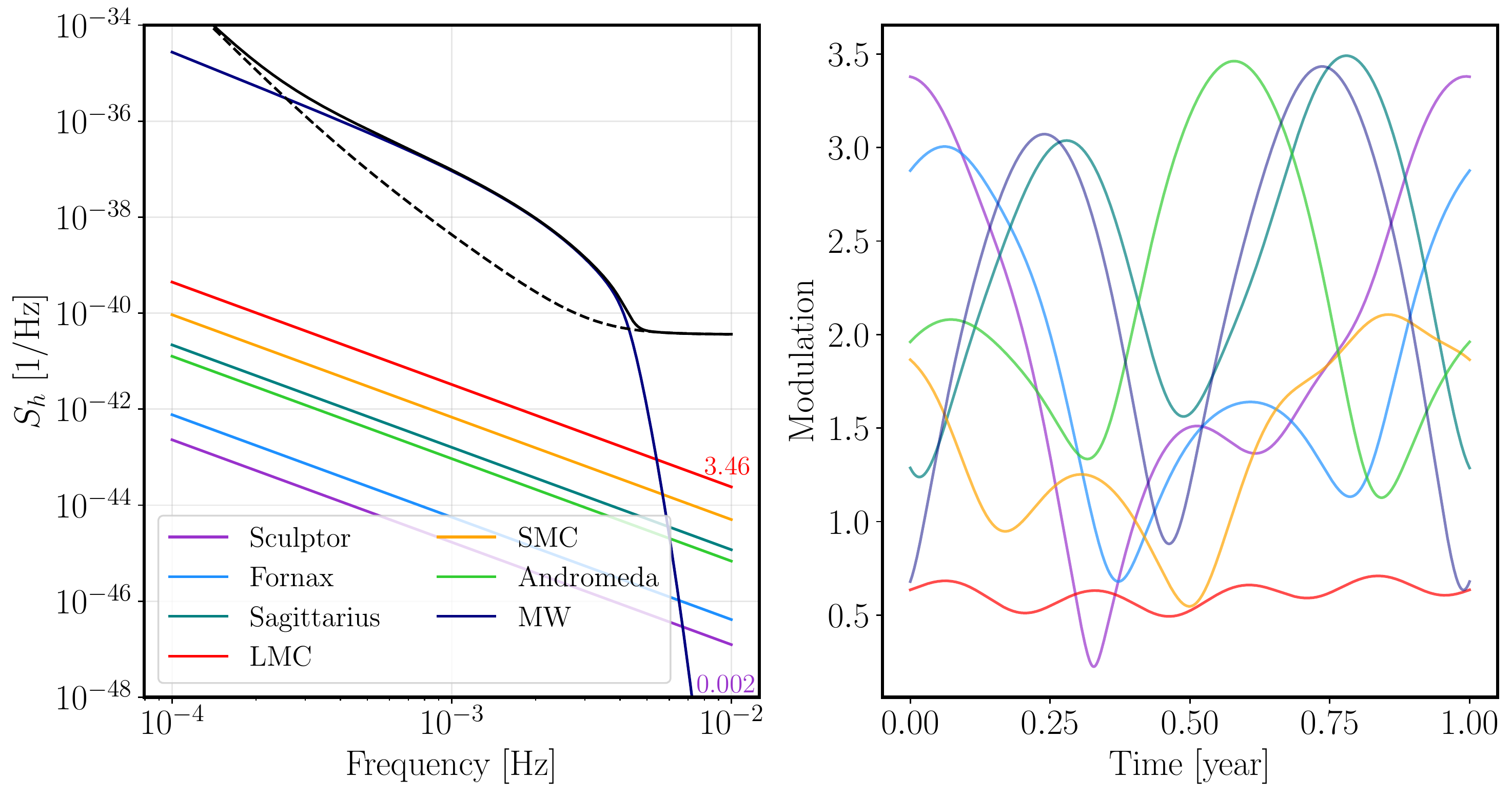}
    \caption{Left panel: The astrophysical SGWB spectrum is computed using a semi-analytical model from Eq.~\eqref{eq:gwb_spec}. The amplitude level mainly depends on the distance and size of the source, in terms of stellar mass (or the number of DWDs within it). The slope of the power-law is $\gamma = -(9+ \alpha)/3$, where $\alpha$, a parameter related to the initial DWD separation distribution, is $\alpha$. The black dashed line refer to the LISA instrumental noise.Right Panel: The modulation in time as seen by LISA during one year for different astrophysical sources, assuming a two-dimensional Gaussian distribution of unresolved DWDs in the sky for each of them. 
    }
    \label{fig:spectra}
\end{figure*}

We now aim to provide a framework to compute the effective astrophysical signal. We present below two simple models to characterize the spectra $S_h$ in Eq.~\eqref{eq:stoch_proc_spectrum} for the Milky Way, its satellites, and other nearby Galaxies.
In Fig.~\ref{fig:spectra} (left panel), we show the spectra for the LMC, and SMC, Sculptor, Fornax, and Sagittarius, as well as Andromeda and the MW foreground.
The Signal-to-Noise Ratio (SNR) values are computed for each source, assuming a LISA mission duration of $4$ years, and are shown for the loudest (LMC, ${\rm SNR}=3.46$) and faintest (Sculptor, ${\rm SNR}=0.002$). 
For simplicity, we show only the upper envelope of each modulated stochastic signal.
The spectra are accompanied by the individual modulations (right panel), as described in Sec.~\ref{sec:lisa_mod}.

In the remainder of this section, we provide further details on the spectral models for each of the sources mentioned before.

~~~\emph{Satellites and nearby Galaxies}~~~
\label{sec:astro_spectrum}
Upon Fourier transforming ${\cal C}$ in Eq.~\eqref{eq:autocorr3} we describe the spectrum of the signal as 
\begin{align}
\label{eq:spectrum}
    S_h(f) &=\frac{N}{8}\int \dd \mathcal{M}_c \int \dd f_s p(\mathcal{M}_c)p(f_s)\times \nonumber\\
    &\times\frac{(G\mathcal{M}_c)^{10/3}}{(c^4 D)^2}\left(\pi f_s\right)^{4/3} \delta(f-f_s)
\end{align}
where for brevity we relabelled $f = \frac{f + f^\prime}{2}$.
The integral in Eq.~\eqref{eq:spectrum} equals the isotropic spectrum amplitude computed in Ref.~\cite{2001astro.ph..8028P} within a spherical shell of radius $D$. It is a convenient approach in that it avoids sky integration, as this has already been accounted for in the computation of the coefficients ($B_n$).

We recast the frequency dependence in the integral in terms
of $x = a/(Kt_0)^{1/4}$.
The variable $x$ represents the separation normalized to that of a binary with time to merger $t_0$, and $a$ is related to $f_s$ through Kepler's law and 
\begin{equation}
    K = \frac{256 G^3}{(5c^5)} (m_1+m_2)m_1m_2.
\end{equation}
Thus, Eq.~\eqref{eq:spectrum} can be rewritten as follows:  
\begin{align}
\label{eq:new_spectrum}
    S_h(f) &=\frac{N}{8}\int \dd \mathcal{M}_c \int \left| \frac{\dd x}{\dd f_s}\right| \dd f_s p(\mathcal{M}_c)p(x(f_s)) \times\nonumber \\
    &\times \frac{(G\mathcal{M}_c)^{10/3}}{(c^4 D)^2}\left(\pi f_s\right)^{4/3} \delta(f-f_s)
\end{align}
The Jacobian in Eq.~\eqref{eq:new_spectrum} reads
\begin{align}
\label{eq:jacobian}
\left| \frac{\dd x}{\dd f_s}\right| = \frac{2}{3}\frac{G^{1/3}}{\pi^{1/3}\Bar{K}^{1/4}}\frac{(m_1 + m_2)^{1/12}}{(m_1m_2)^{1/4}}f_s^{-5/3},
\end{align}
and upon substitution in Eq.~\eqref{eq:new_spectrum} and integrating with the Dirac delta, we get
\begin{align}
    S_h(f) 
    &\propto\int \dd \mathcal{M}_c \mathcal{M}_c^{35/12}f^{-1/3}p(\mathcal{M}_c)p(x(f_s = f))
\end{align}
This is a convenient parametrization of the integral, as it matches closely the one adopted in previous population studies (see, e.g. Ref.~\cite{2012ApJ...751..143M,2018MNRAS.476.2584M,2022MNRAS.511.5936K}). 
In particular, we adopt the following parametrization
\begin{equation}
    p(x) \propto \begin{cases}
        x^{4 +\alpha}\left[(1+x^{-4})^{\frac{\alpha +1}{4}}-1\right], & \mbox{if } \alpha \neq -1 \\
        x^3\ln{\left(1+x^{-4}\right)}, & \mbox{if } \alpha = -1       
    \end{cases}.
\end{equation}
where $\alpha$ is the power-law slope describing the separation distribution at DWD formation and the star formation is assumed to be constant (for detailed derivation see Ref.~\cite{2012ApJ...751..143M}). 
In our analysis, we will consider $\alpha = -1.3$ following results obtained in Ref.~\cite{2018MNRAS.476.2584M}, based on two complementary large, multi-epoch, spectroscopic samples: the Sloan Digital Sky Survey (SDSS), and the Supernova Ia Progenitor surveY (SPY). 
When $\alpha\leq -1$, in the limit of $x \ll 1$, $p(x)$ can be approximated by $x^{4+\alpha}$, so that the final model for $S_h(f)$ is
\begin{equation}
\label{eq:gwb_spec}
        S_h(f) \propto f^{-(9 + 2\alpha)/3}\int \dd \mathcal{M}_c \mathcal{M}_c^{5(3-\alpha)/12}p(\mathcal{M}_c).
\end{equation}

Eq.~\eqref{eq:gwb_spec} provides a semi-analytical power-law describing the SGWB spectrum for a generic MW satellite with constant star formation. This may be an optimistic assumption for lower-mass satellites, which typically stop forming stars early, but it represents a good approximation for satellites like the SMC and LMC that continue forming stars. Its amplitude depends on the integral in the chirp mass, which we perform numerically from the distribution on $m_1$ and $m_2$. We assume that $m_1$ follows the mass distribution of single white dwarfs.
Following Ref.~\cite{2022MNRAS.511.5936K}, we consider a univariate Gaussian mixture with three components for $m_1$ with weights $w = (0.66, 0.24, 0.1)$, means $\mu = (0.55, 0.72, 0.3) \rm M_{\odot}$, and standard deviations $\sigma = (0.07, 0.3, 0.078) M_{\odot}$, as measured in Ref.~\cite{2015MNRAS.446.4078K} based on large spectroscopic sample. Observations show that at the main-sequence stage, the secondary stars follow a mass-ratio distribution that is approximately flat, rather than the same mass distribution as the primary (e.g., \cite{2017ApJS..230...15M}) 
Therefore, in our model we draw $m_2$ from a flat distribution between $0.15 \rm M_\odot$, the minimum mass of observed extremely-low-mass (ELM) white dwarfs, and  $m_1$. 
For a small number of cases in which we get  $m_1 < 0.25 M_\odot$\footnote{Given such low masses, ELM white dwarfs require more than a Hubble time to form via single stellar evolution, meaning their masses can only be explained by binary interactions. Therefore, they are treated differently in our modeling.}, i.e., falling into the ELM category, 
we draw  $m_2$ from the range $[0.2,1.2] M_\odot$ with equal probability. Note that in the latter case, the definitions of $m_1$ and $m_2$ are swapped. 

The amplitudes of the SGWB at $f = 10^{-3.5}\rm Hz$ are reported for different satellites in Table~\ref{tab:par}. 
Therefore, a satellite spectrum with assumed mass distribution has only two free parameters: we reparameterize them with the power-spectrum amplitude at a reference frequency $A(f = 10^{-3.5}\rm Hz)$ and its slope $\gamma = -(9 +3\alpha)/3$.
Notably, when $\alpha = -1$, the slope simplifies to $-7/3$, which is the well-known value associated with a SGWB originating solely from inspiral signals.

~~~\emph{Milky Way}~~~
The overall spectral shape of the  Galactic foreground  is influenced by the astrophysical characteristics of the actual DWD population and is predicted to exhibit two main features: a low-frequency tail and a high-frequency knee around a few millihertz. Assuming only that the orbital evolution depends solely on GW emission, the low-frequency tail is expected to follow a power-law with index equal to $-7/3$ expressed as PSD. A variety of ways of parameterizing the shape of the Galactic foreground exist ~\cite{2021PhRvD.104d3019K,2019JCAP...11..017C,2021JCAP...01..059F}.
We decide to adopt the phenomenological model introduced in Ref.~\cite{2021PhRvD.104d3019K}
\begin{align}
    \label{eq:mw_spectrum}
    S_h(f)\!=\! \frac{A}{2}f^{-7/3}e^{-(f/f_1)^{\alpha_{\rm MW}}}\!\left(1 \!+\! \tanh\!\left(\frac{f_{\rm knee}-f}{f_2}\right)\!\right).
\end{align}
The overall amplitude $A$ characterizes the population of DWDs. 
The frequency $f_2$ and the parameter $\alpha_{\rm MW}$ are expected to be properties of the population and the SNR threshold for individual sources detection considered. 
The remaining parameters, $f_1$ and $f_{\rm knee}$, are shown to be strongly dependent on the LISA observation time. 
Indeed, the longer the observation period, the more DWDs are detected and therefore removed from the unresolved population.

\section{Towards parameter estimation}\label{sec:lisa-data}
In this study, we focus exclusively on the parameter estimation of satellites' and MW SGWBs together with the instrumental noise.  We proceed with the assumption that non-stochastic GWs sources have been subtracted perfectly. Additionally, we consider the absence of data gaps, resulting in the generation of an idealized residual dataset. Throughout the analysis, we consider 4 years of observation time of LISA. Data are expressed in terms of time-delay interferometric (TDI) variables~\cite{2005LRR.....8....4T}. Specifically, we work with the Fourier transform of the noise-orthogonal channels, denoted as $\Tilde{d} = (\Tilde{A}, \Tilde{E}, \Tilde{T})$.

While zero correlation between TDI channels is not assured in a realistic setup~\cite{2023PhRvD.107l3531H}, we have chosen to neglect off-diagonal terms in our analysis. 
We defer the extension of our analysis to incorporate this effect to future work.
Additionally, the T variable is often considered a null channel in LISA, meaning it is effectively insensitive to gravitational waves. However, in a realistic instrumental setup, this cannot be guaranteed across the entire sensitivity band~\cite{2023PhRvD.107l3531H}. 
Therefore, we exclude the T channel from our analysis.

\subsection{LISA Instrumental Noise}
As previously mentioned, we model the datastreams as a superposition of independent stochastic processes: the astrophysical signal and the instrumental noise. Specifically, LISA's instrumental noise is considered stationary and Gaussian. Assuming that the primary noise source, laser frequency noise, has been perfectly removed through the use of TDI, we include only secondary noise sources in the data, namely the test mass (TM) noise and the optical metrology system (OMS) noise. The respective power spectral densities are:
\begin{align}
\label{eq:noise_tm}
        S_{n}^{\rm tm}\left(f\right) &= P_{\rm tm}\left[1+\left(\frac{0.4 \rm mHz}{f} \right)^2\right] \left[1+\left(\frac{f}{8 \rm mHz} \right)^4\right] \nonumber\\
        &\times \left( \frac{1}{2\pi fc}\right)^2 \left( \rm m^2/ \rm s^3\right),
\end{align}
and
\begin{equation}
\label{eq:noise_oms}
    S_{n}^{\rm oms}\!\left(f\right)\! =\!P_{\rm oms}\!\left[1\!+\!\left(\frac{2 \rm mHz}{f} \right)^4\right]\!\!
    \left( \frac{2\pi f}{c}\right)^2\!\!\!\!\left(\rm m^2 \!\cdot \rm s\right)
\end{equation}
with $P_{\rm tm} = (3 \times 10^{-15})^2$ and $P_{\rm   oms} = (15 \times 10^{-12})^2$~\cite{LISA2018}. The two noise component are propagated in TDI channels through transfer functions~\cite{2023PhRvD.108h2004Q}.

\subsection{Likelihood}
Inference is performed simultaneously on signal and instrumental noise. We construct joint posteriors on parameters $\boldsymbol{\theta}$
\begin{equation}
    p(\boldsymbol{\theta}|\Tilde{d}) \propto \mathcal{L}(\Tilde{d}|\boldsymbol{\theta})\pi(\boldsymbol{\theta})
\end{equation}
through stochastic sampling of the likelihood $\mathcal{L}(\Tilde{d}|\boldsymbol{\theta})$ under chosen priors $\pi(\boldsymbol{\theta})$. 
To do so, we use  \textsc{Balrog}, a large codebase for simulation and inference on LISA signals.
In this study, we use it alongside a nested sampling algorithm as implemented in \textsc{Nessai}~\cite{2021PhRvD.103j3006W} to obtain each posterior and evidence. We assume that data are distributed according to the Whittle likelihood. In the most general case, we include in the likelihood the contribution from the Galactic foreground,  the satellite SGWB and the instrumental noise: 
\begin{align}
&\log\mathcal{L}(\tilde{d}|\boldsymbol{\theta} = \{\boldsymbol{\theta}_{\rm MW},\boldsymbol{\theta}_{\rm sat},\boldsymbol{\theta}_{\rm n}\}) = \rm{const}  + \nonumber\\ 
&- \sum_{i = A,E} \frac{1}{2}\log(\rm det \left[\Sigma_{\rm d}\right]_i) +\frac{1}{2}\tilde{d}_i^T\left[\Sigma_{\rm d}\right]_i^{-1}\tilde{d}_i\\
\left[\Sigma_{\rm d}\right]_i &= \left(\Sigma_{\rm MW}(\boldsymbol{\theta}_{\rm MW}) + \Sigma_{\rm sat}(\boldsymbol{\theta}_{\rm sat}) + \Sigma_{\rm n}(\boldsymbol{\theta}_{\rm n})\right)_i
\end{align}
where $\Sigma_{\rm MW}$ and $\Sigma_{\rm sat}$  correspond to the MW and satellite covariance matrices as described in Sec.~\ref{sec:cyc}, and $\left[\Sigma_{\rm d}\right]_i$ denotes the covariance matrix of the data in the $i$-th channel. $\Sigma_n$ is the diagonal noise covariance matrix with PSD described in Eqs.~\eqref{eq:noise_tm} and~\eqref{eq:noise_oms}. The parameters for each process are: 
\begin{itemize}
    \item $\boldsymbol{\theta}_{\rm MW} =  \left\{{\cal A}_{\rm MW}, \alpha, f_{\rm knee}, f_2, f_1, \lambda, \sin \beta, \sigma_1, \sigma_2, \psi\right\}$ ;
    \item $\boldsymbol{\theta}_{\rm sat} = \left\{{\cal A}_{\rm sat}, \gamma, \lambda, \sin\beta, \sigma\right\}$;
    \item $\boldsymbol{\theta}_{n} = \left\{ {\cal P}_{\rm tm}, {\cal P}_{\rm oms} \right\}$.
\end{itemize}
Calligraphic letters indicate the $\log_{10}$ of the corresponding quantity, e.g. $\mathcal{A}_{\rm MW} = \log_{10} A_{\rm MW}$. 
We choose uniform priors for each parameter listed.

We compare results obtained under both cyclostationary and stationary inference. 
To achieve this, we use the log-Bayes factor:
\begin{equation}
    \log_{10}\mathcal{B}^{\rm cyclo}_{\rm stat} = \log_{10}\mathcal{Z}_{\rm cyclo} - \log_{10}\mathcal{Z}_{\rm stat},
    \label{eq:bayes_factor}
\end{equation}
where 
\begin{equation}
    \mathcal{Z} = \displaystyle \int \dd \boldsymbol{\theta} \mathcal{L}(\Tilde{d}|\boldsymbol{\theta})\pi(\boldsymbol{\theta})
\end{equation}
is the Bayesian evidence.

\section{Results}\label{sec:res}

We now present our inference results. 
First, we examine LISA's capability to observe sources in the presence of instrumental noise alone. 
Next, we assess the impact of including the MW foreground in both simulated data and inference model.
In addition, we investigate the detectability of MW largest nearby Galaxy, Andromeda.

\begin{table*}[ht]\centering
    \centering
    \renewcommand{\arraystretch}{1.3}
    \begin{tabular}{c|c|c|c|c|c|c}
    \hline
    System & $\lambda$ & $\beta$ & Sky Area $[\rm deg^2]$ & $D$ $[\rm kpc]$&  Stellar Mass $[\rm M_{\odot}]$ & $\log_{10}A(f = 10^{-3.5}\rm Hz)$ \\
    \hline
    Andromeda & $27.8^\circ$ & $33.3^\circ$ & 3.11& 765.0& $\sim10^{10}$& -41.96\\
    Fornax & $22.5 ^\circ$& $-46.9^\circ$& 0.17& 139.0& $2.0 \times 10^7$& -43.18\\
    LMC & $-47.7^\circ$& $-85.4^\circ$& 77& 50.0& $1.5\times10^9$&-40.42\\
    SMC & $-47.9 ^\circ$& $-64.6^\circ$& 13& 60.6& $4.6\times10^8$& -41.098\\
    Sagittarius & $-78.1^\circ$& $-7.6^\circ$& 37& 26.7& $2.1\times10^7$&-41.73\\
    Sculptor & $-1.6^\circ$& $-36.5^\circ$& 0.076& 86.0& $2.3\times10^6$&-43.704\\
    \hline
    \end{tabular}
    \caption{Property of MW satellites and Andromeda 
    Mass, distance, and sky location are taken from  Refs.~\cite{2012AJ....144....4M,2014MNRAS.445..881C,2019ARA&A..57..375S}. The last column represents the logarithmic amplitude of the SGWB computed as defined in Sec.~\ref{sec:astro_spectrum}.}
    \label{tab:par}
\end{table*}

\begin{table*}[t]
    \centering
    \renewcommand{\arraystretch}{1.3}
    \begin{tabular}{c|c|c|c}
    \hline
    Data & Model & Parameters & References \\ 
    \hline
    SMC  & Cyclo & $(A, \gamma, \sigma)$ & Fig.~\ref{fig:posterior_sgwb} (Panel 1, Noise in Fig.~\ref{fig:noise_rec}) \\
    LMC  & Cyclo & $(A, \gamma, \sigma)$ & Fig.~\ref{fig:posterior_sgwb} (Panel 2, Noise in Fig.~\ref{fig:noise_rec}) \\
    Andromeda  & Cyclo & $(A, \gamma, \sigma)$ & Fig.~\ref{fig:posterior_sgwb} (Panel 3, Noise in Fig.~\ref{fig:noise_rec})\\
    Sagittarius  & Cyclo & $(A, \gamma, \sigma)$ & Fig.~\ref{fig:posterior_sgwb} (Panel 4, Noise in Fig.~\ref{fig:noise_rec}) \\
    MW + SMC  & Cyclo + Cyclo & $(\alpha, A, f_2, \psi, \sigma_1, \sigma_2) + (A, \gamma, \sigma)$ & Fig.~\ref{fig:smc_corner} \\
    MW + LMC  & Cyclo + Cyclo & $(\alpha, A, f_2, \psi, \sigma_1, \sigma_2) + (A, \gamma, \sigma)$& Fig.~\ref{fig:lmc_corner} \\
    MW + Andromeda  & Cyclo + Cyclo & $(\alpha, A, f_2, \psi, \sigma_1, \sigma_2) + (A, \gamma, \sigma)$ & Fig.~\ref{fig:and_corner} \\
    MW + Sagittarius  & Cyclo + Cyclo & $(\alpha, A, f_2, \psi, \sigma_1, \sigma_2) + (A, \gamma, \sigma)$ & Fig.~\ref{fig:sag_corner} \\
    LMC (realistic)  & Cyclo & $(A, \gamma, \sigma)$ & Fig.~\ref{fig:lmc_cyclovsstat} \\
    LMC (realistic)  & Stat & $(A, \gamma)$ & Fig.~\ref{fig:lmc_cyclovsstat} \\
    MW + LMC-like& Cyclo + Cyclo & $(\alpha, A, f_2, \psi, \sigma_1, \sigma_2) + (A, \gamma, \lambda, \sin\beta, \sigma)$ & Fig.~\ref{fig:mw_corner}\\
    \hline
    \end{tabular}
    \caption{The text summarizes the parameter estimation as follows: The first column details the sources injected into the data, with each dataset also including LISA instrumental noise. The second column describes the model used to represent the signals. The third column lists the parameters employed in the inference process, while the final column references the figure depicting the corresponding posterior distribution.\label{tab:PE}}
\end{table*}

\subsection{Milky Way satellites and Andromeda}

The characteristic parameters  of the galaxies and satellites we investigate are reported in Table~\ref{tab:par}. Table~\ref{tab:PE}
provides a summary of the parameter estimation discussed in this section and the following ones.
\subsubsection{Mock Satellite Data}
\begin{figure*}
    \centering
    \includegraphics[width=0.5\columnwidth]{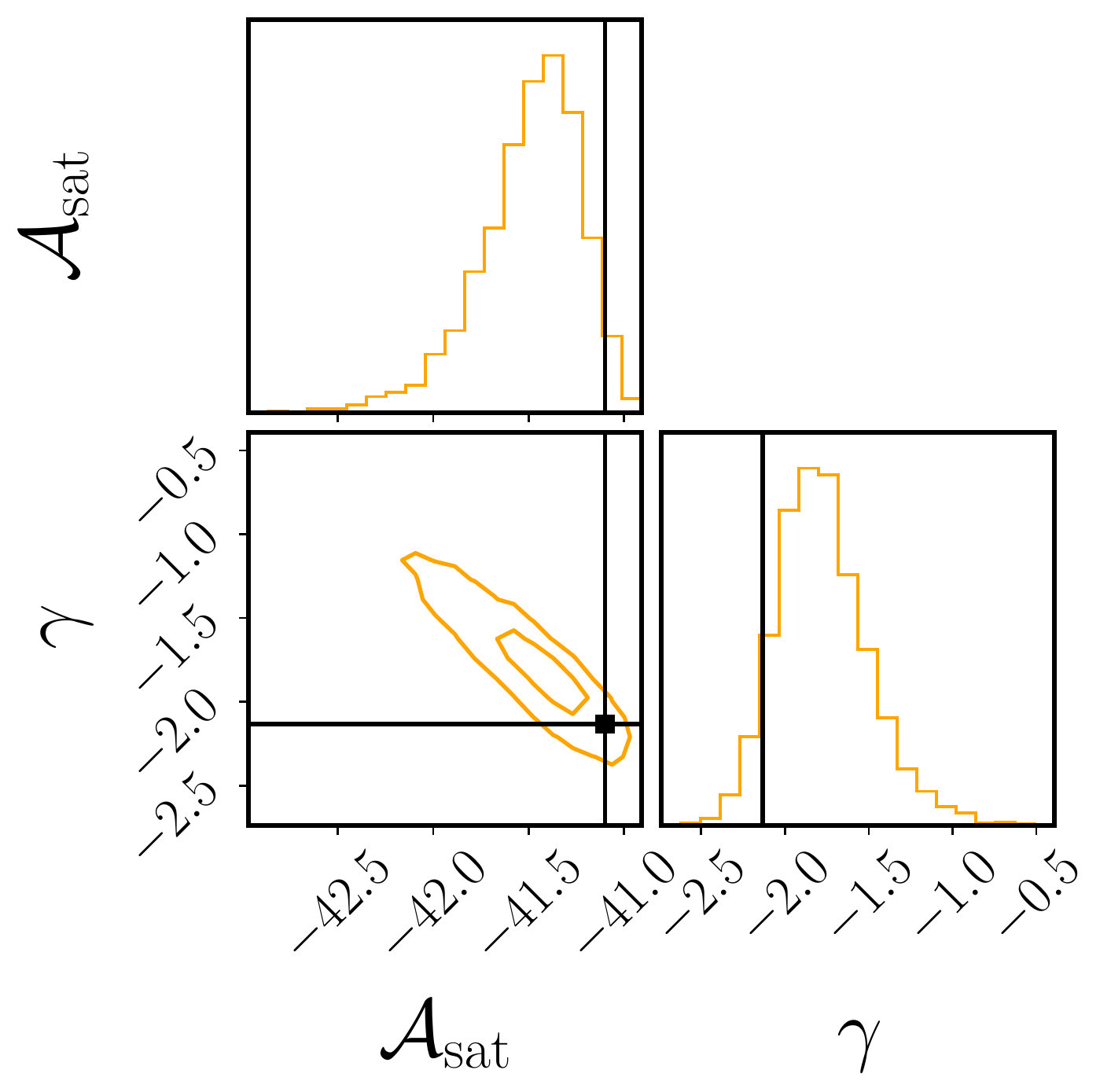}
    \includegraphics[width=0.5\columnwidth]{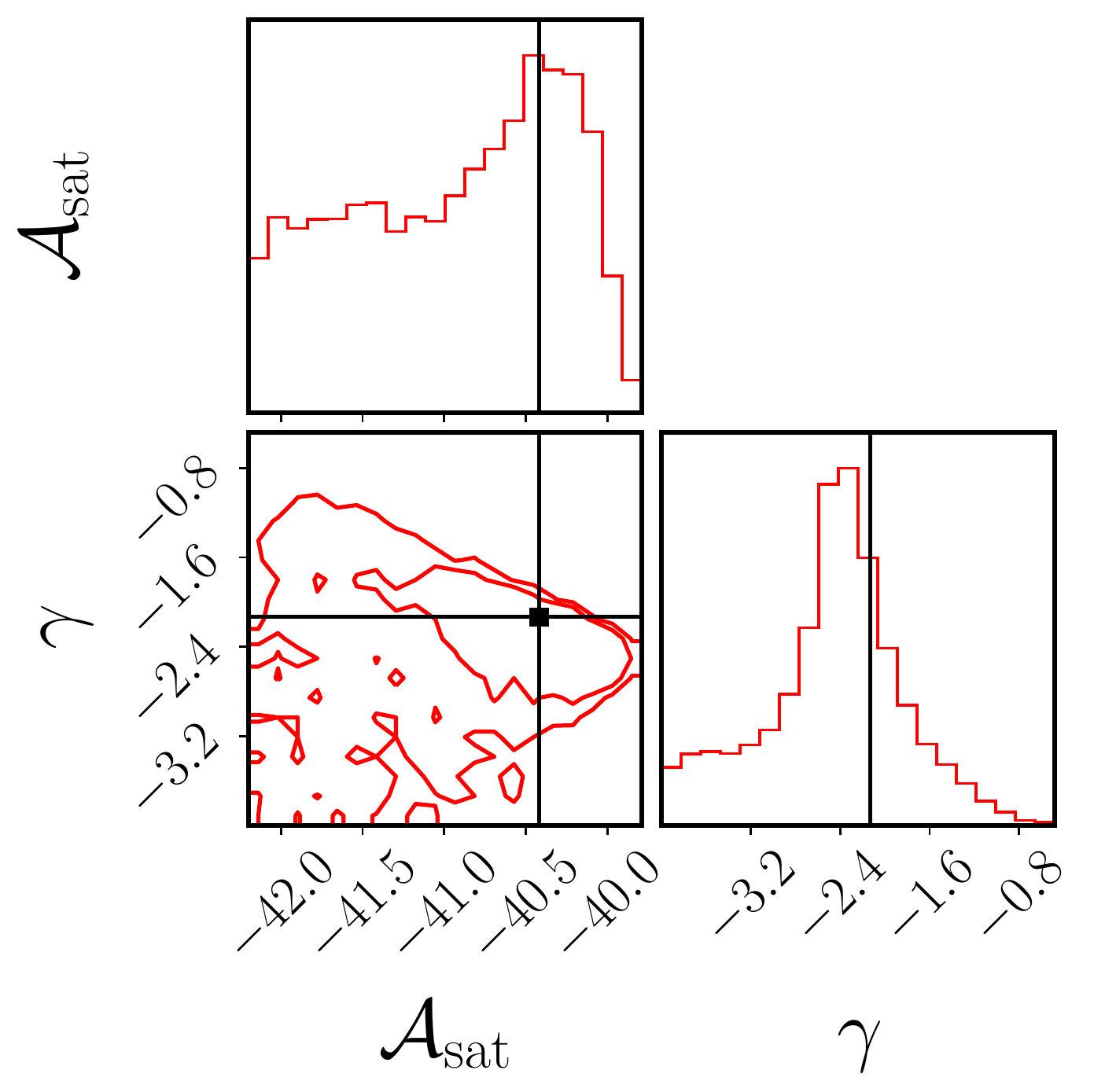}
    \includegraphics[width=0.5\columnwidth]{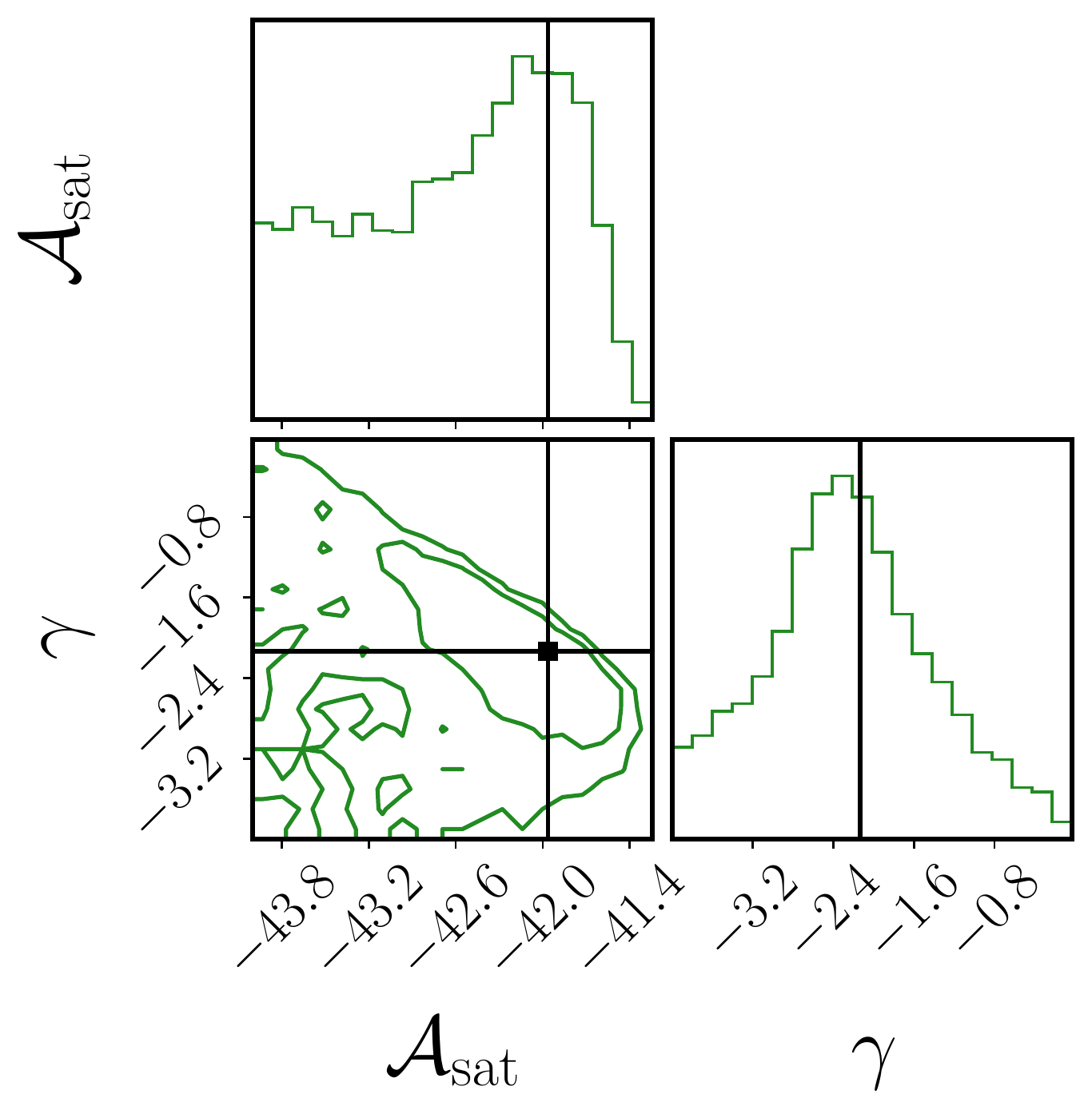}
    \includegraphics[width=0.5\columnwidth]{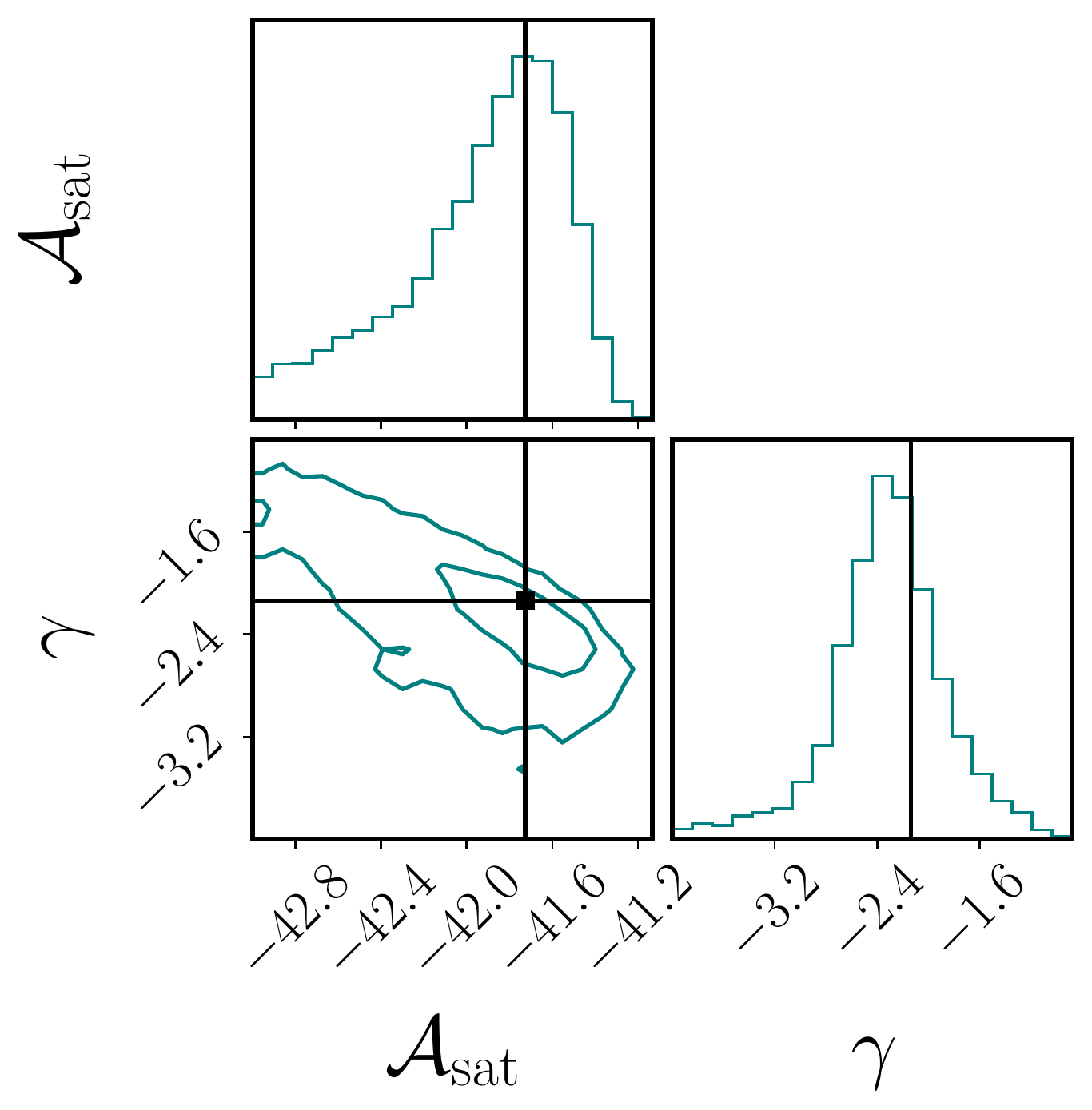}
    \caption{Marginalized probability distribution of the amplitude and slope SGWB parameters for SMC, LMC, Andromeda, and Sagittarius (from left to right). In this analysis, the Galactic foreground is excluded from both the injection and parameter estimation stages. Additionally, we assume that the sky locations of the satellites are known throughout the inference process.}
    \label{fig:posterior_sgwb}
\end{figure*}

\begin{figure}
    \includegraphics[width=1.0\columnwidth]{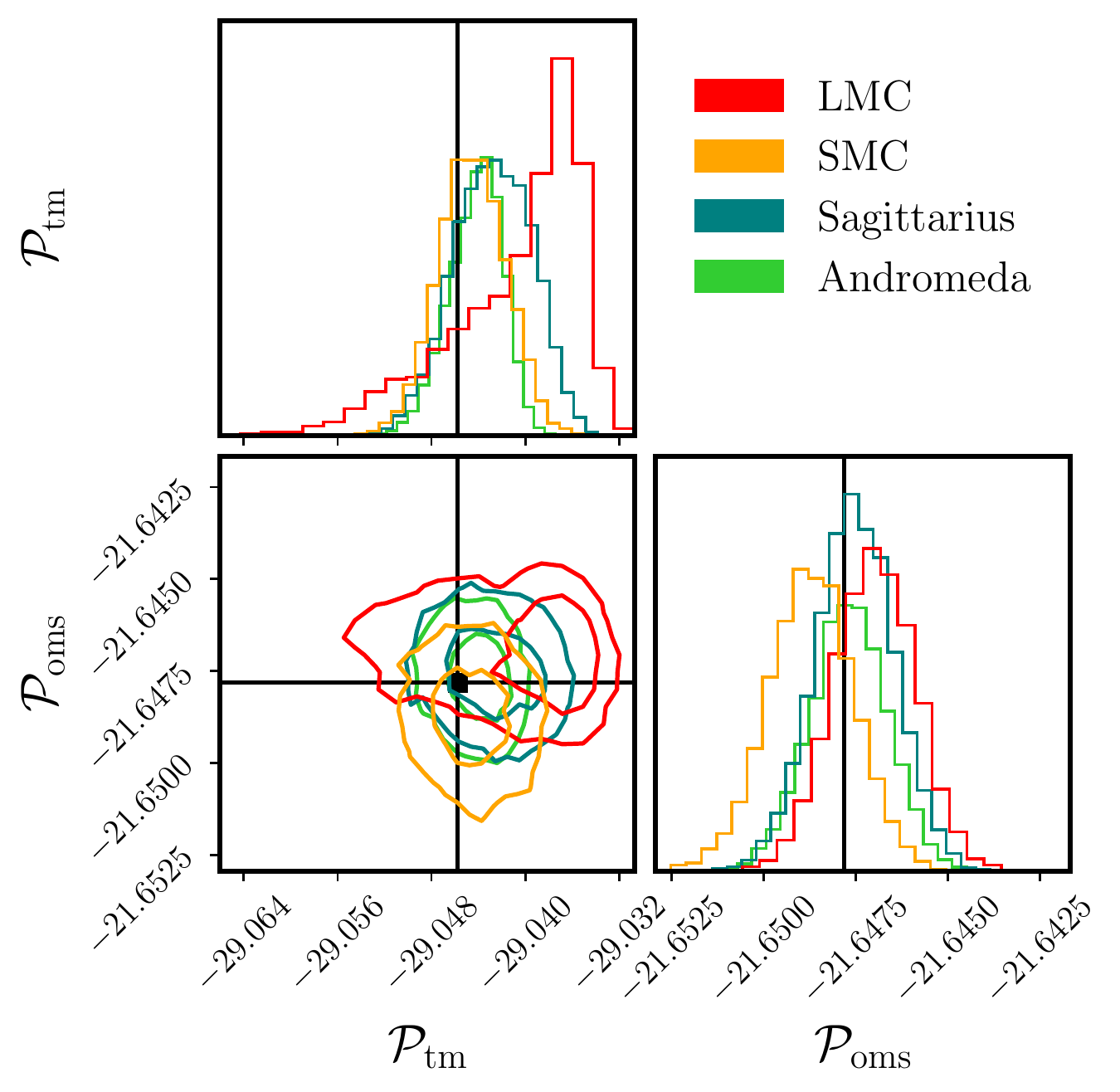}
    \caption{Marginalized posterior probability of TM and OMS noise amplitudes. Colors correspond to the source SGWB injected with noise. The presence of the SGWB influence the reconstruction of the noise, in particular the TM component results to be overestimated. The Galactic foreground is not included in these analyses.}
    \label{fig:noise_rec}
\end{figure}

To determine which MW satellite might be observable by LISA, we first study its detectability in presence of instrumental noise only, excluding the MW foreground from the data and parameter estimation. Specifically, we inject idealized satellites one by one, with their spectra and modulation described in Sec.~\ref{sec:cyc}, along with their corresponding amplitude, sky location, and angular radius.  

Initially, we also fit the extrinsic parameters of the SGWB source, specifically the size and sky location. The results indicate that we can clearly recover both the spectrum amplitude and slope only for the SMC. Regarding other parameters, the $\sin\beta$ value closely matches the true value, while the ecliptic longitude appears difficult to estimate accurately. 
This inaccuracy is likely due to the SMC's proximity to the South Pole, where longitudes partial degeneracy become stronger approaches it. 
The size $\sigma$ is not constrained due to its mild influence on the modulation with respect to the latitude, as explained in Sec.~\ref{sec:lisa_mod}. 

From this initial analysis, we report a central finding: 
the ability to detect a satellite is influenced by the interplay between the astrophysical spectrum and the modulation effects in reconstructing this type of SGWB. From an astrophysical point of view, the LMC is favored due to its mass and distance, which makes it more probable to detect individual DWD binaries~\cite{2020ApJ...894L..15R}. 
However, from the perspective of SGWB detection, the SMC proves to be easier to detect due to its location.

Despite the SGWB signal from satellites still being unknown, these objects have already been observed through electromagnetic emission. Therefore, in what follows, we assume perfect prior knowledge of their locations to determine if this improves the spectrum parameter estimation. 
By doing so, the spectral reconstruction improves for LMC, Sagittarius and Andromeda. The SGWB amplitude is well constrained only for SMC. The marginalized posterior probability of their slope and amplitude are shown in Fig.~\ref{fig:posterior_sgwb}. Although LISA is more sensitive to the sky positions of Fornax and Sculptor compared to the LMC or SMC, the spectra from these satellites are too faint to be detected, even if their positions (hence their signal modulation) are known. 

In Fig.~\ref{fig:noise_rec}, we plot the marginalized posterior distribution for the two noise parameters. The presence of an SGWB, even if it cannot be measured with high precision (Fig.~\ref{fig:posterior_sgwb}), can still introduce biases in the instrumental noise reconstruction. The TM amplitude results are  slightly overestimated. This effect is particularly evident in the case of the LMC. Indeed, the covariance matrix has small off-diagonal terms due to the faint modulation of the LMC, making the TM noise and the SGWB spectrum more degenerate at low-frequency along the main diagonal.

Next, we consider the case where the MW foreground is included in the data, alongside the instrumental noise and each satellite signal. The Galactic signal is generated using the spectrum given by Eq.~\eqref{eq:mw_spectrum} with the optimal fit parameters found in Ref.~\cite{2021PhRvD.104d3019K}, considering 4 years of LISA observation time and an SNR threshold of 7 for DWD detection. 
For the modulation, we adopt the same strategy used by Ref.~\cite{Buscicchio2024}, where it is generated with $\lambda = -93.16^\circ$, $\beta = -5.53^\circ$, $\sigma_1 = 0.042$, $\sigma_2 = 0.138$, and $\psi = -56.72^\circ
$. 
These parameters result from the best fit of the interpolation between a rotated two-dimensional Gaussian distribution and a realistic unresolved MW catalog of DWDs.
This approach is an approximation that only partially captures the more complex structure of the Milky Way (as depicted in Fig.5 of Ref.~\cite{Buscicchio2024}). 
A more sophisticated  description of the MW time-variability is deferred to future work, where we employ a mixture of distributions to better reconstruct the global structure of the MW.

In this scenario, as expected, the detection of MW satellites is highly compromised. The posterior distribution for SMC, LMC, Andromeda, and Sagittarius are presented in the Appendix ~\ref{sec:appendix}. Using informative priors on sky location (both for the satellite and MW), we are able to put an upper limit on the amplitude of the SGWB spectrum, while the slope is in general easier to constrain. 
Nonetheless, a good reconstruction is guaranteed for both the MW foreground and the instrumental noise.
In particular, for the Milky Way, we infer only $A$, $f_2$, and $\alpha$, which are the most relevant parameters from an astrophysical point of view. 
We fix the remaining parameters to their injected values due to computational limitations. By doing so, the nested sampling algorithm will converge more quickly without introducing significant biases in the analysis. 

We emphasize that, unlike what is observed for the satellites, we are indeed able to recover the MW size. This is due to the fact that the MW angular size is sufficiently large that modulation effects become sensitive to fluctuations around the true value in the prior parameter space. 
This aligns with our expectations from Fig.~\ref{fig:prior_harm}.

\subsubsection{Realistic LMC \label{subsubsec:realistic-lmc}}
After applying our cyclostationary model to ideal scenarios, we aim to test it in the presence of realistic datasets.
Therefore, we inject a realization of the SGWB generated from a realistic catalog of individual DWDs. We consider the case of the LMC, since its source catalog has been already used in previous work \cite{2024MNRAS.530..844K,korol_2024_10854469}. 

Similarly to the previous section, we fix the satellite's position in our model. The results align well with those obtained using mock data. We successfully constrain the spectral parameters, particularly the slope. However, the satellite's size has a minimal impact on the model, making it difficult to effectively constrain this parameter. We report the posterior distribution in Fig.~\ref{fig:lmc_cyclovsstat} in Appendix~\ref{sec:appendix}.
We also infer the signal under the assumption of stationarity, modeling the background signal using a power-law. It’s important to note that the amplitude here directly refers to the TDI domains. Additionally, since the T channel is consistently excluded from our analysis, we simplify the model by using a single amplitude parameter associated with both the A and E channels.
To compare the amplitude between two models, we divide the TDI amplitude in the stationary case by the zero order Fourier coefficient, computed using the real satellite parameters (size and sky position). 
It is interesting to note that in both models, the slope of the background is consistent with our simplified theoretical model described in Sec.~\ref{sec:astro_spectrum}. 

We compute the Bayes factor as described in Eq.~\eqref{eq:bayes_factor}. We find that $\log_{10}\mathcal{B}^{\rm{cyclo}}_{\rm{stat}} \lesssim -0.5$, which does not indicate a strong evidence in favor of one of the two hypothesis. This is due to the faintness of the signal modulation. The off-diagonal elements in the covariance matrix are not significant enough to help distinguish the signal from the noise in the main diagonal. As a result, the stationary model can extract the same amount of information. Instead, for the Galactic signal, exploiting the off-diagonal elements will be crucial. However, as previously mentioned, a detailed analysis of realistic data for the MW foreground will be addressed in future work.  

\subsection{Discovering hidden satellite}
\label{subsubsec:newsatellite}
\begin{figure}
    \includegraphics[width=\columnwidth]{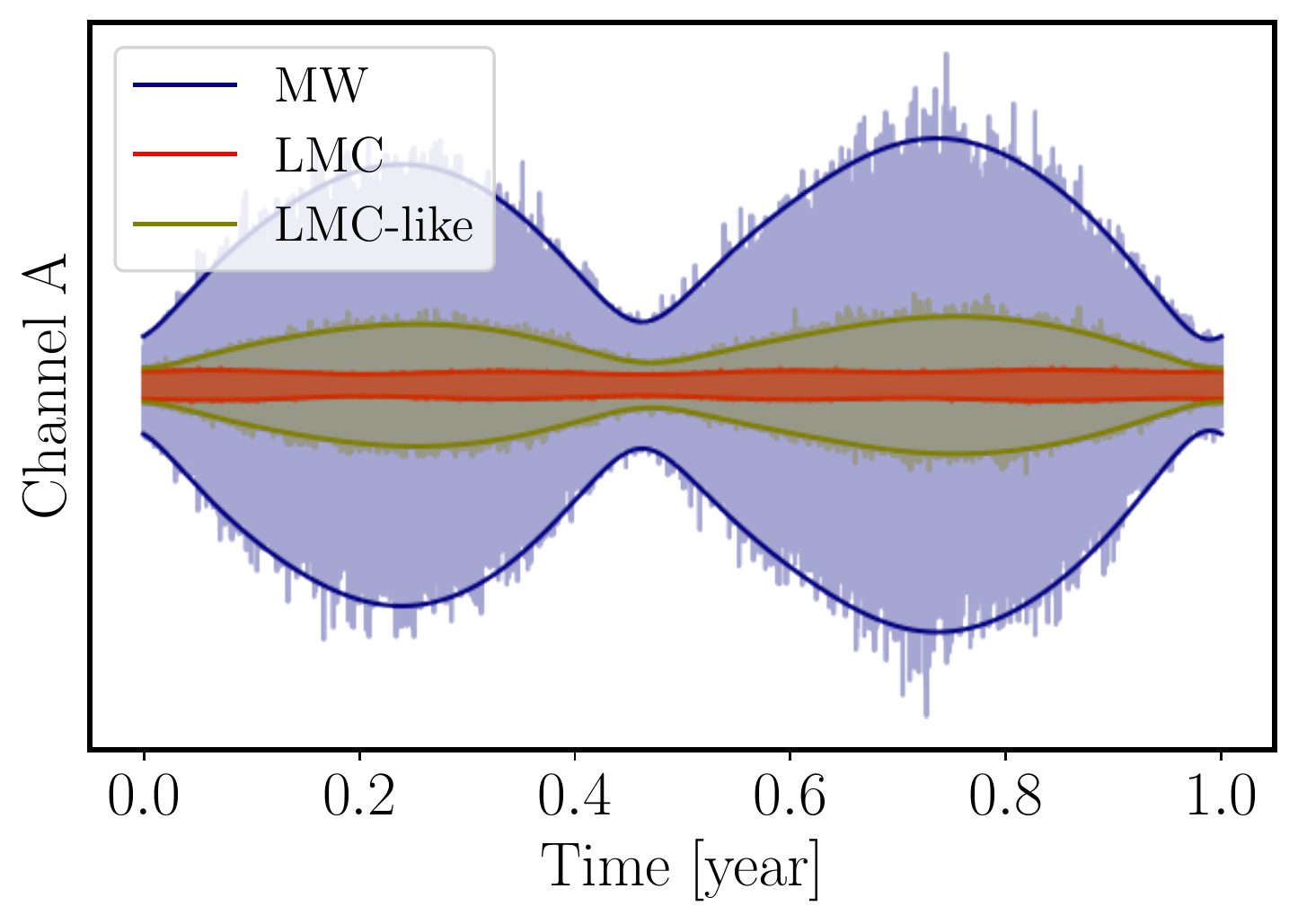}
    \includegraphics[width=\columnwidth]{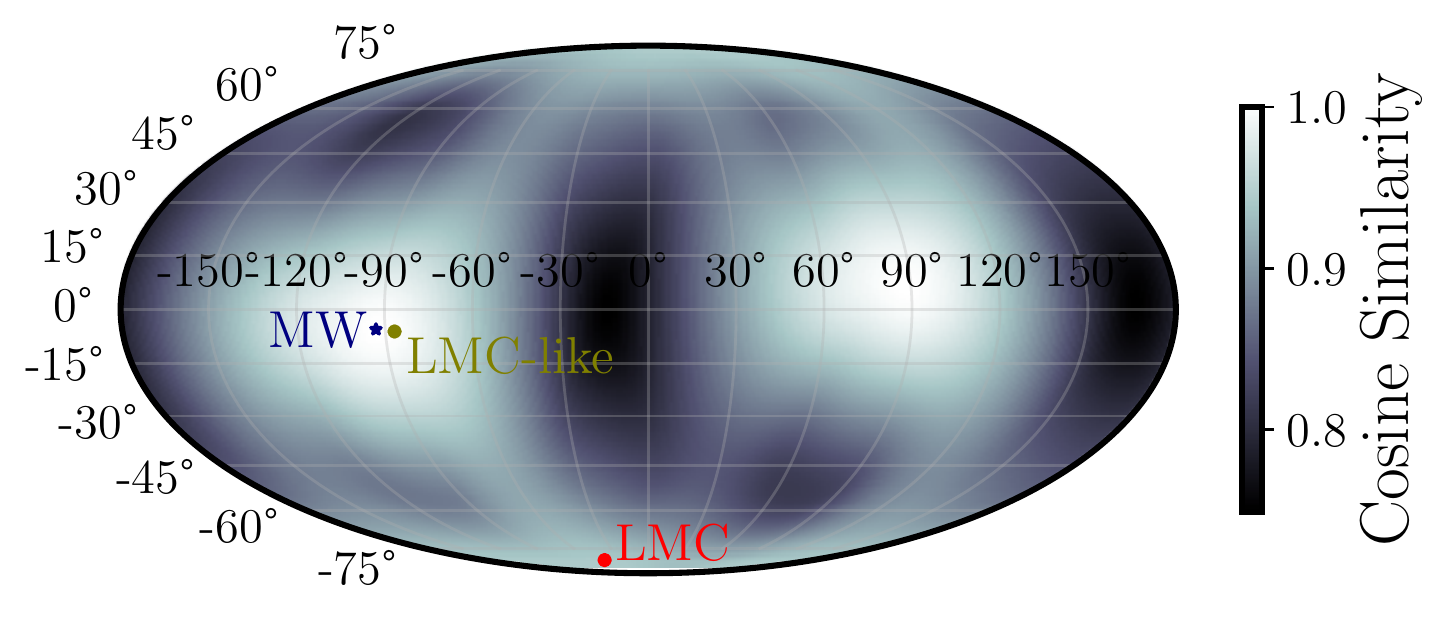}
    \caption{Top Panel: Simulated signals for MW (teal) and LMC (red), and shifted LMC within the Galactic disk (light blue) field of view over the course of one year. The corresponding modulations are represented as envelopes superimposed on the top and bottom of the signals. Bottom Panel: Cosine similarity map between the Milky Way (MW) and the LMC-like satellite modulations. The three dots represent: the center of the MW (navy), the true position of the LMC (red), and the hypothetical position of the LMC-like satellite in the Galactic disk (gold).}
    \label{fig:time_signal}
\end{figure}

Unlike electromagnetic emission, which is obstructed by dust and gas in the Zone of Avoidance, GWs are not hindered by such obstructions.
Thus, with LISA, we have the potential to observe beyond the Galactic plane. According to  Ref.~\cite{2020ApJ...894L..15R}, detecting an excess of $\sim 100$ DWD binaries within the Galactic disk would be considered a statistically significant overdensity. In an ideal scenario, if a satellite were located behind the MW disk, we could attempt to detect its SGWB. For this reason, we inject an LMC like SGWB (i.e., with the same slope and amplitude), in a region close to the Galactic center. For simplicity, we consider the same $~\sigma$. In Fig.~\ref{fig:time_signal}, we display the simulated signals for the MW (navy), LMC (red), and the hypothetical LMC behind the MW (green olive). We observe that moving the LMC away from the pole enhances LISA’s sensitivity to the corresponding SGWB. This adjustment causes the modulation of the signal to become similar to the MW one. Consequently, this can lead to a degenerate covariance matrix structure between the two processes, potentially resulting in inaccurate signal reconstruction or, in the worst case, failure to detect the SGWB.     
However, this does not occur. As shown in Fig.~\ref{fig:mw_corner}, we successfully disentangle the SGWB from both the LISA instrumental noise and the MW foreground. 
Additionally, we reanalyze the same dataset assuming LMC-like SGWB as stationary. In this case, we not only obtain a smaller evidence ($\log_{10}\mathcal{B}^{\rm{cyclo}}_{\rm{stat}} \gtrsim 1000$) but also introduce biases in both the foreground and background parameters. This finding reinforces our earlier point regarding the realistic LMC dataset analysis: as the signal becomes more modulated, the importance of cyclostationarity increases for accurately resolving the background, leading to stronger evidence. However, it is important to note that, contrary to the realistic LMC analysis, the dataset is generated under ideal conditions. 

\section{Conclusion}\label{sec:conc}
In this work, we develop a method to exploit the cyclostationarity of SGWB, originating from unresolved DWD binaries in the MW satellite galaxies or nearby galaxies, for its recovery and parameter estimation.   
These signals are characterized by their astrophysical spectrum and the time-domain modulation induced the LISA response to the anisotropic foregrounds (back) due to the constellation annual motion.

Based on a simple, physically-motivated argument, we approximate the astrophysical spectrum with a power-law. 
We instead employ an analytical model for the time-variability, which we incorporate into a Bayesian, frequency-domain, inference scheme.

LISA detection capability strongly depends on the interplay between the spectrum loudness and LISA's sensitivity to the position of each source. 
We show how parameter estimation improves with perfect knowledge of the sky-location, which is a realistic assumption since these objects are already studied through electromagnetic emission.
However, as expected, including the MW foreground in the analysis leads to a significant reduction in LISA's sensitivity to the SGWB.
We apply the cyclostationary model also the realistic data from the LMC. We find similar results to the one obtained with mock data. We are able to put constraint on the SGWB spectral shape. However, given the faintness of the signal and its modulation, the cyclostationary model is not strongly preferred in respect to the stationary approach. 

Additionally, we study LISA's capability to discover new satellites through their GW emission. We demonstrate that LISA can detect an LMC-like satellite located behind the Galactic disk, which would be largely invisible through electromagnetic radiation. In contrast to the realistic LMC, cyclostationarity is strongly favored due to enhanced modulation. Further analysis of a realistic realization of this signal is required to confirm this significant result.

In our analysis, we use an approximate model to account for the quasi-stationarity of the MW foreground. In future work, we plan to develop a more realistic model that captures the time variability arising from a complex distribution of sky sources, extending beyond a simple bivariate Gaussian. We anticipate that the loud intensity of the foreground will allow us to capture anisotropies from various source distributions across the sky.
A more detailed characterization of the Galactic foreground will not only refine the description of other GW signals but also enable us to explore the structure of the Milky Way through gravitational waves.
Additionally, no instrumental artifacts or extragalactic SGWB components are expected to exhibit cyclostationary behavior. However, in realistic scenarios, the characterization of the covariance matrix may be compromised by systematic effects from the resolution of other sources. This is a crucial aspect that warrants further investigation in future analyses.

\begin{acknowledgments}

The authors thank N.~Karnesis, A.~Vecchio, and C.~J.~Moore for useful insights and fruitful comments on this study.
R.B. acknowledges support through the Italian Space Agency grant \emph{Phase A activity for LISA mission, Agreement n. 2017--29--H.0}, by the MUR Grant ``Progetto Dipartimenti di Eccellenza 2023-2027'' (BiCoQ), and by the ICSC National Research Center funded by NextGenerationEU.\@
A.K. acknowledges support of the UK Space Agency grant, no. ST/V002813/1.
A.S. acknowledges financial support provided under the European Union’s H2020 ERC Consolidator Grant ``Binary Massive Black Hole Astrophysics'' (B Massive, Grant Agreement: 818691).
Computational work was performed at University of Birmingham BlueBEAR High Performance Computing facility,
at Bicocca's Akatsuki cluster (B Massive funded), at CINECA with allocations through EuroHPC Benchmark access call grant EUHPC-B03-24, and at Google Cloud through award No.~GCP19980904.

\textit{Software}:
We acknowledge usage of 
\textsc{Mathematica}~\cite{Mathematica} 
and of the following 
\textsc{Python}~\cite{10.5555/1593511} 
packages for modeling, analysis, post-processing, and production of results throughout:
\textsc{Nessai}~\cite{2021PhRvD.103j3006W},
\textsc{matplotlib}~\cite{2007CSE.....9...90H},
\textsc{numpy}~\cite{2020Natur.585..357H},
\textsc{scipy}~\cite{2020NatMe..17..261V}.
\end{acknowledgments}
\clearpage
\appendix
\onecolumngrid

\section{Posterior distributions}
\label{sec:appendix}
\begin{figure}[ht!]   \includegraphics[width=\columnwidth]{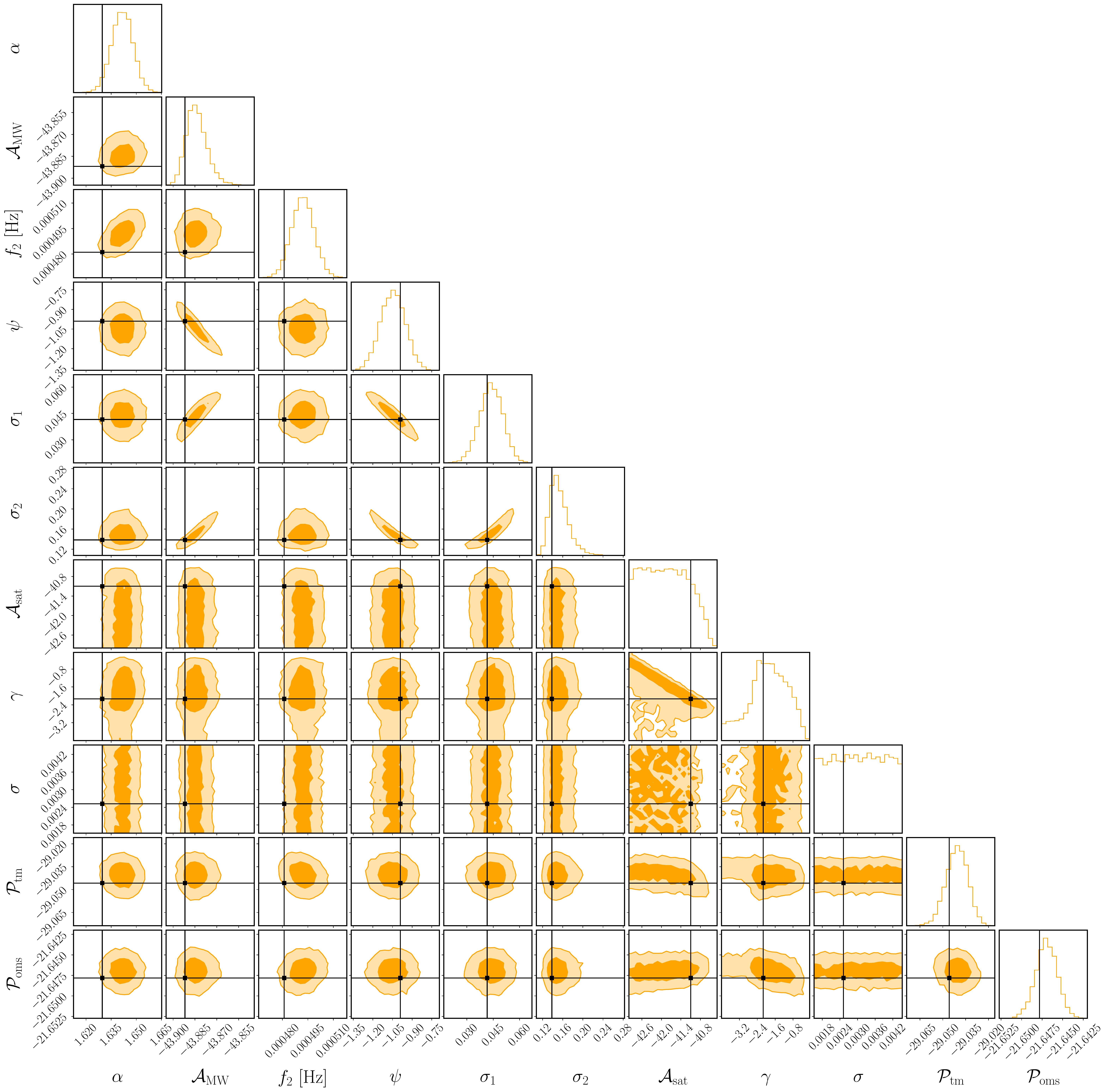}
    \caption{Posterior distribution on SMC's SGWB in the presence of noise and MW foreground. Darker (lighter) shaded areas denote $90\%$ ($50\%$) credible regions and black lines indicate the true values.}
    \label{fig:smc_corner}
\end{figure}

\begin{figure}[ht!]   \includegraphics[width=\columnwidth]{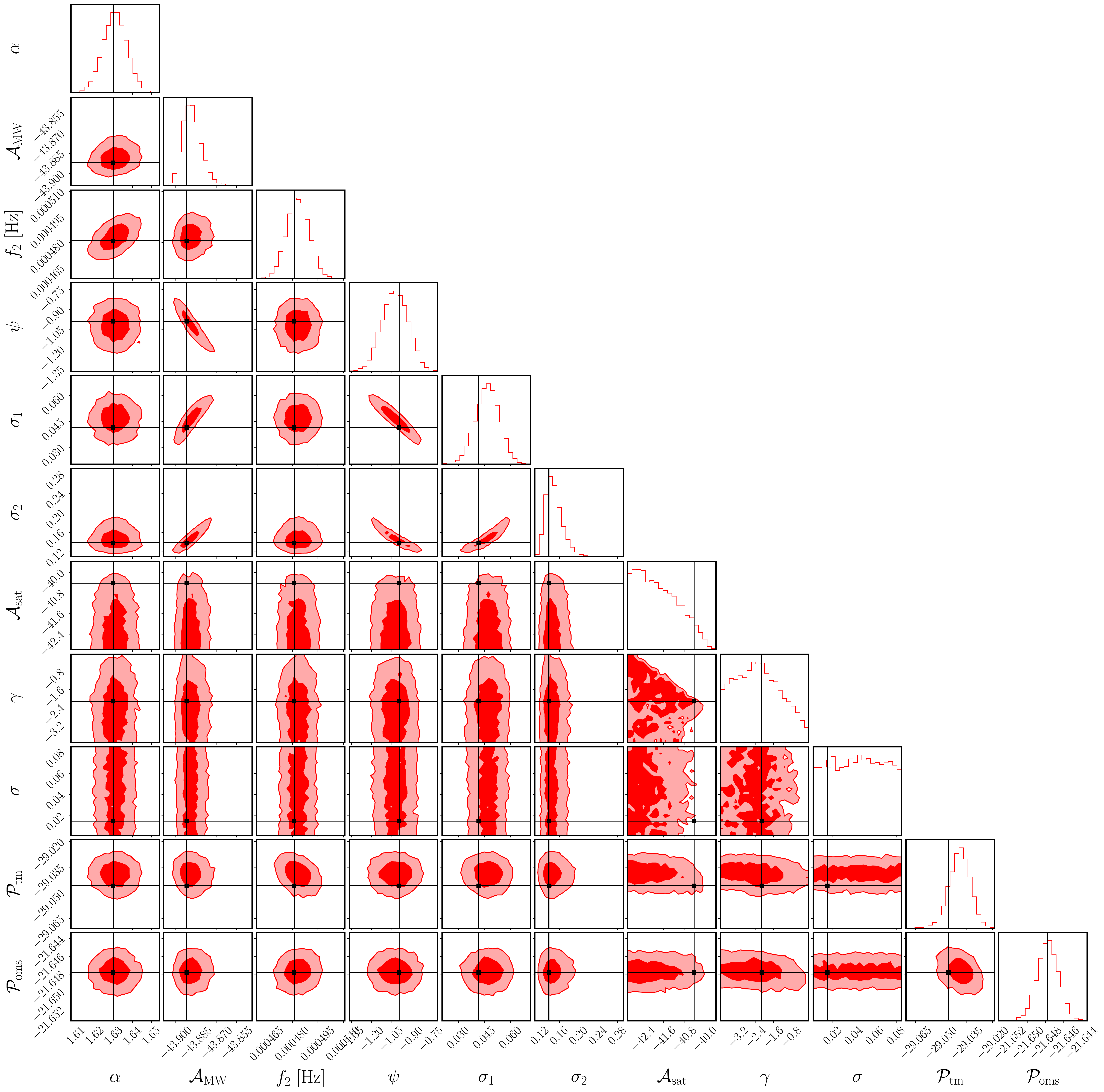}
    \caption{Posterior distribution on LMC's SGWB in presence of noise and MW foreground. Darker (lighter) shaded areas denote $90\%$ ($50\%$) credible regions and black lines indicate the true values.}
    \label{fig:lmc_corner}
\end{figure}

\begin{figure}[ht!]   \includegraphics[width=\columnwidth]{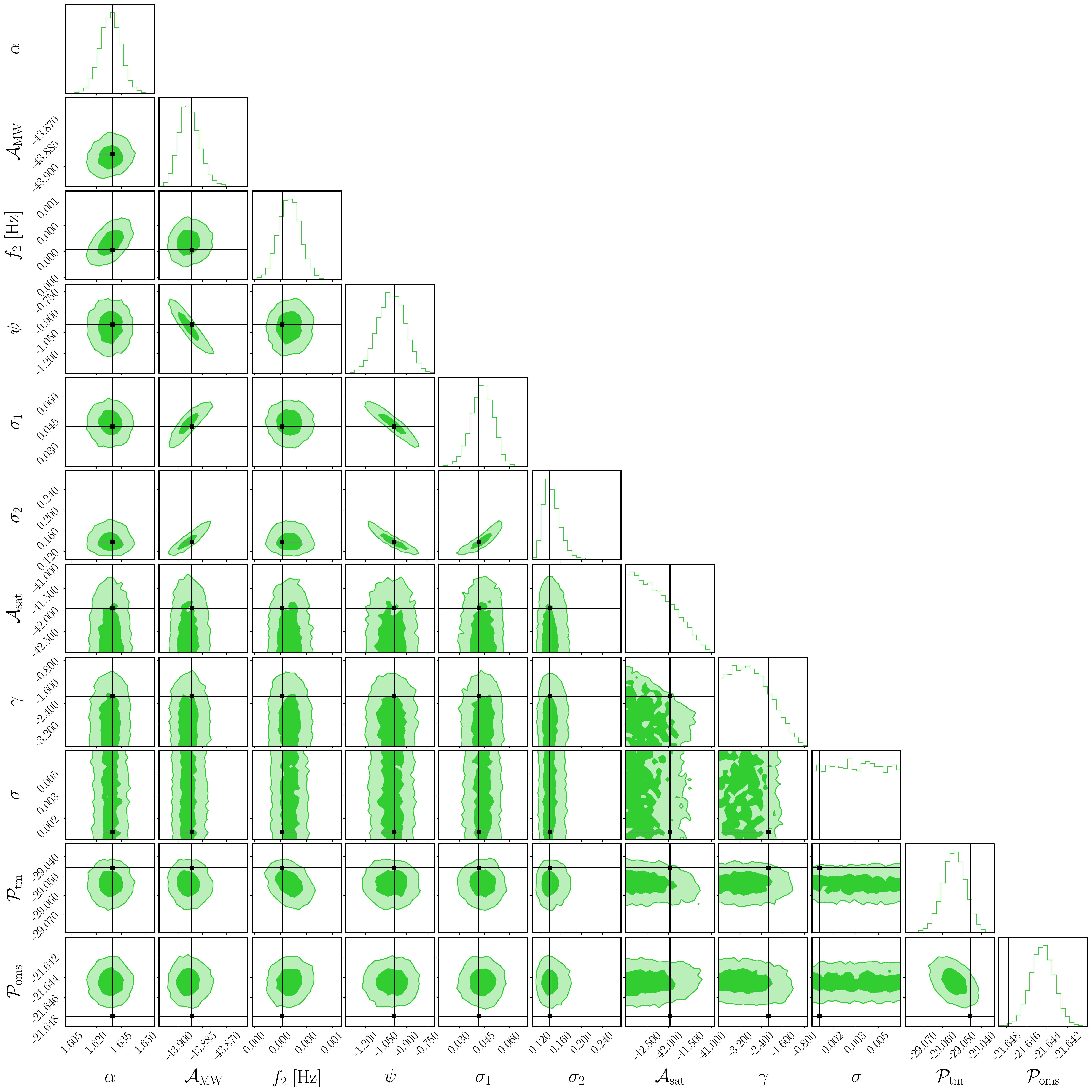}
    \caption{Posterior distribution on Andromeda's SGWB in presence of noise and MW foreground. Darker (lighter) shaded areas denote $90\%$ ($50\%$) credible regions and black lines indicate the true values.}
    \label{fig:and_corner}
\end{figure}

\begin{figure}[ht!]   \includegraphics[width=\columnwidth]{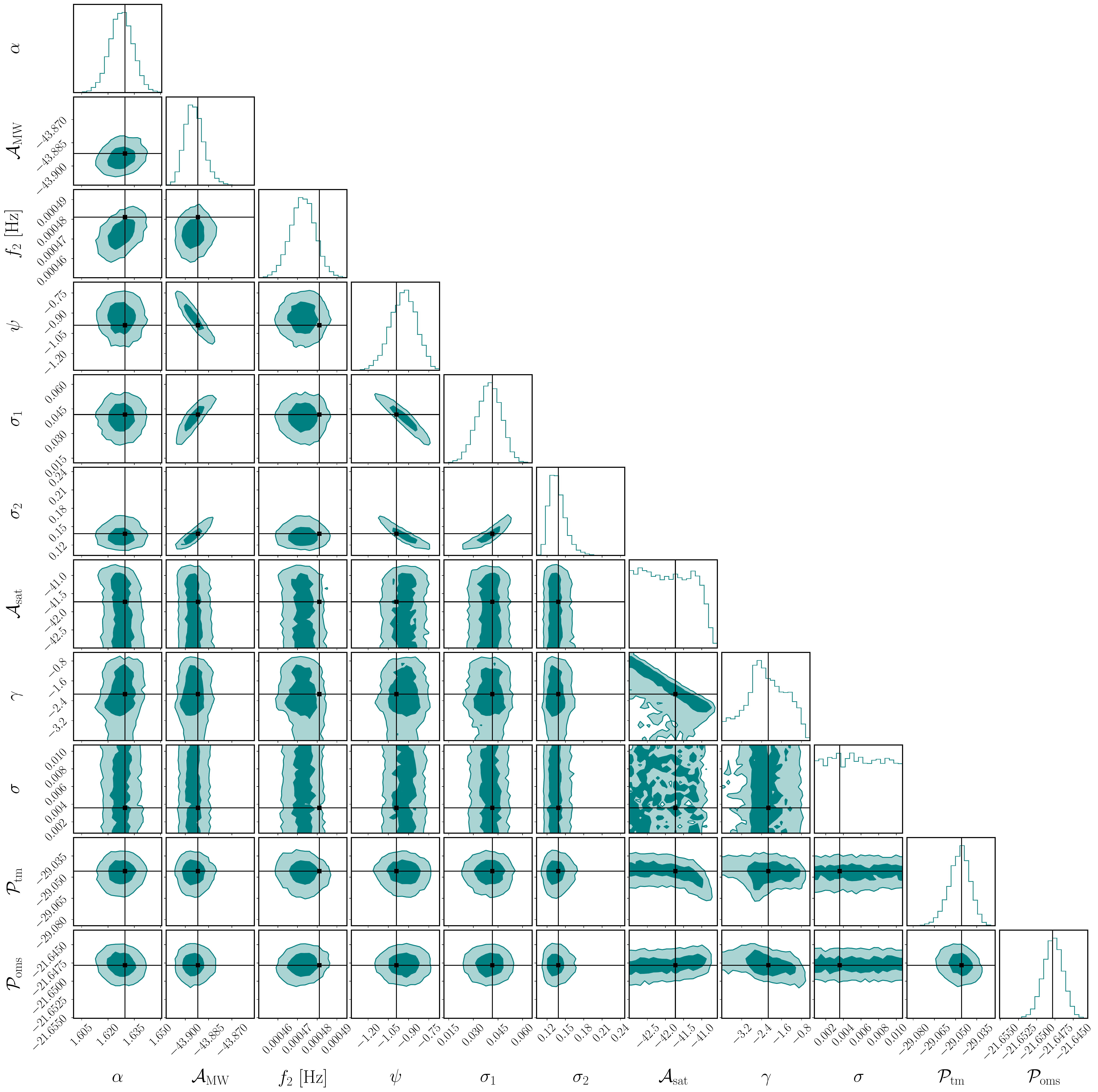}
    \caption{Posterior distribution on Sagittarius' SGWB in presence of noise and MW foreground. Darker (lighter) shaded areas denote $90\%$ ($50\%$) credible regions and black lines indicate the true values.}
    \label{fig:sag_corner}
\end{figure}

\begin{figure}[ht!]   \includegraphics[width=0.9\columnwidth]{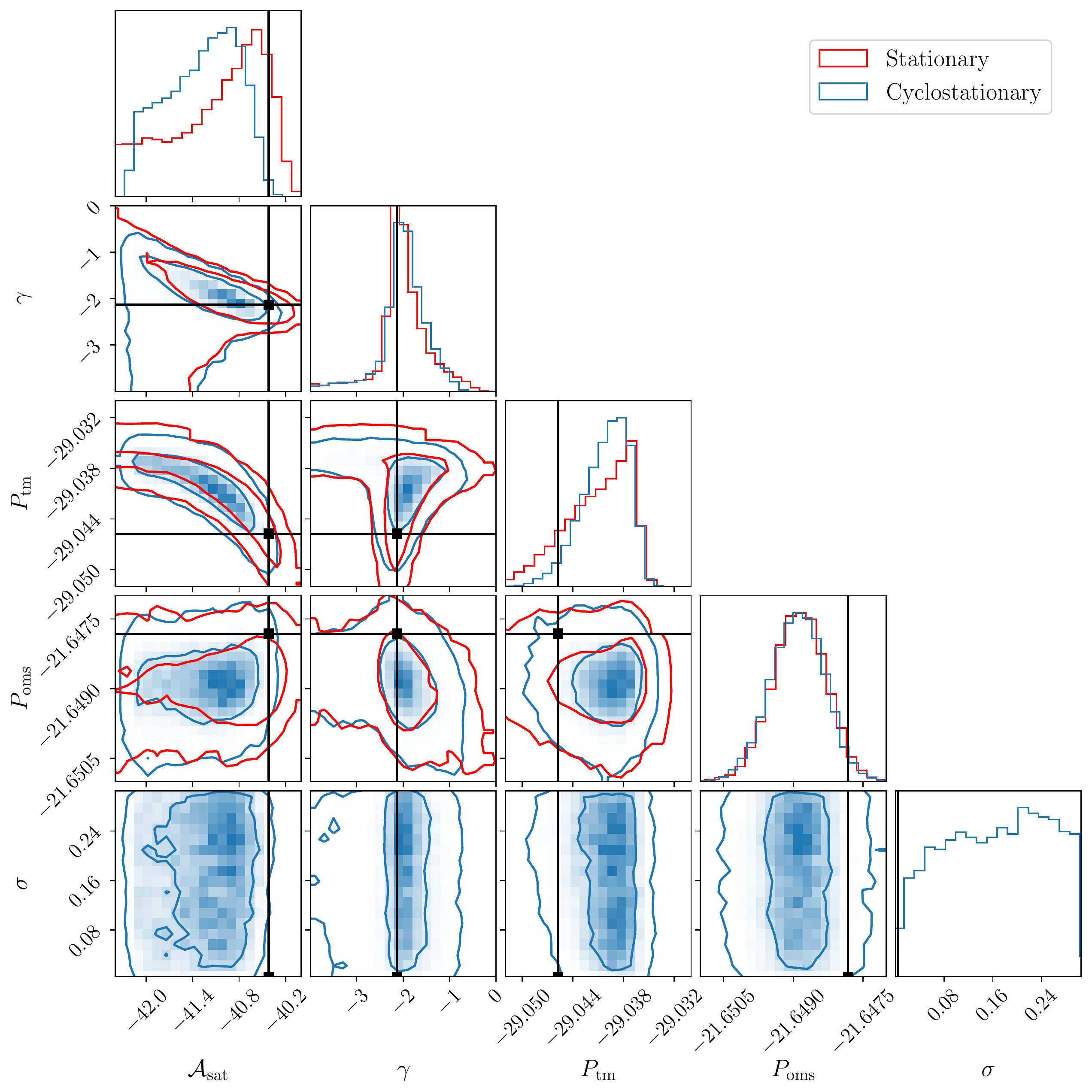}
    \caption{The posterior distribution for noise and LMC signal's parameters, under the stationary (red) and cyclostationary (blue) hypotheses, as discussed in Sec.~\ref{subsubsec:realistic-lmc}. 
    The SGWB data are generated using a realistic catalog of unresolved sources. 
    In the stationary model, the SGWB amplitude is directly modelled in TDI domain, whereas in the cyclostationary model, it refers to the astrophysical strain. 
    To facilitate comparison between the two approaches, the amplitude of the stationary model in the plot has been rescaled using the modulation coefficient of the 0-th harmonic. Darker (lighter) shaded areas denote $90\%$ ($50\%$) credible regions and black lines indicate the true values.}
    \label{fig:lmc_cyclovsstat}
\end{figure}

\begin{figure}[ht!]   \includegraphics[width=\columnwidth]{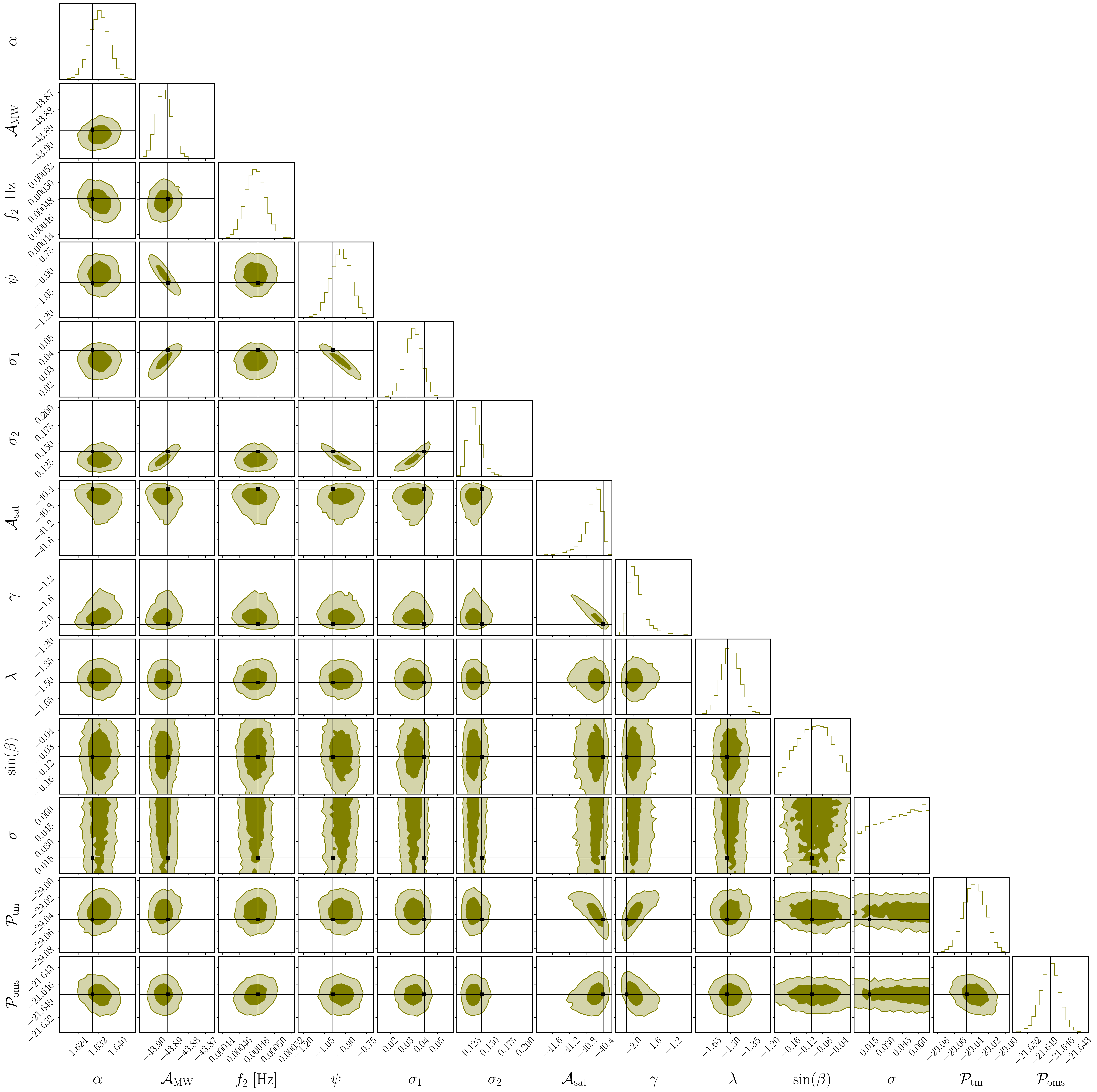}
    \caption{Posterior distribution of MW foreground, noise, and an LMC-like mock signal emitted from the Galactic disk. The plot is discussed in Sec.~\ref{subsubsec:newsatellite}. Darker (lighter) shaded areas denote $90\%$ ($50\%$) credible regions and black lines indicate the true values.}
    \label{fig:mw_corner}
\end{figure}

We present the results of 
parameter estimations 
in Sec.~\ref{sec:res}. 
Corner plots in~\Cref{fig:smc_corner,fig:lmc_corner,fig:and_corner,fig:sag_corner},
illustrate the posterior distributions 
analyzing mock data containing
instrumental noise, SGWB, and Galactic Foreground.
Both the MW and SMC signals are modeled 
as cyclostationary, following 
the formalism in Sec.~\ref{sec:cyc}. 
The true positions of the source centers are 
assumed known. 
In general, the MW signal is well constrained both in terms of its spectrum $(\alpha, \mathcal{A}_{\rm{MW}}, f_2)$ and modulation parameters $(\psi, \sigma_1, \sigma_2)$. 
The presence of the MW 
let us set an upper bound only on the SGWB amplitude, 
while the spectral slope 
remains accurately constrained. 
Fig.~\ref{fig:lmc_cyclovsstat} refers to realistic LMC data and compares the stationary and cyclostationary models. 
Figure \ref{fig:mw_corner} shows the reconstruction of the SGWB from an hypothetical satellite
located 
behind the Galactic disk, 
in 
the presence of the MW foreground and instrumental noise.

\clearpage
\twocolumngrid

\nocite{*}
\clearpage
\bibliographystyle{apsrev4-2}
\bibliography{main}
\end{document}